\documentclass[12pt]{iopart}
\usepackage{iopams}  
\usepackage[latin1]{inputenc}
\usepackage[]{graphicx}
\usepackage{color}

\begin{document} 

\title[Relaxational mode and 
phonon-coupling effects  
in PMN]{Quasi-Elastic Scattering, Random Fields and 
phonon-coupling effects in PbMg$_{1/3}$Nb$_{2/3}$O$_3$}

\author{S N Gvasaliya$^{1,3}$\footnote[3]{To
whom correspondence should be addressed (severian.gvasaliya@psi.ch)}
 B  Roessli$^1$ R A Cowley$^2$ P
Hubert$^1$ and S G Lushnikov$^3$
}

\address{$^1$Laboratory for Neutron Scattering ETHZ \& Paul Scherrer
Institut, CH-5232 Villigen PSI, Switzerland}

\address{$^2$Clarendon Laboratory, Oxford University, Parks Road, Oxford
OX1 3PU, UK}

\address{$^3$Ioffe Physical Technical Institute, 26 Politekhnicheskaya,
194021, St. Petersburg, Russia}

\begin{abstract}
The low-energy part of the vibration spectrum in PbMg$_{1/3}$Nb$_{2/3}$O$_3$
(PMN) relaxor ferroelectric has been studied by neutron scattering above 
and below the Burns temperature, T$_d$. 
The transverse acoustic and the lowest transverse optic phonons 
are strongly coupled and we have obtained a model for this coupling. 
We observe that the lowest optic branch is always underdamped. A 
resolution-limited central peak and quasi-elastic scattering appear 
in the vicinity of the Burns temperature. It is shown that it is 
unlikely that the quasi-elastic scattering originates from the combined 
effects of coupling between TA and TO phonons with an increase of the 
damping of the TO phonon below T$_d$. The quasi-elastic scattering has a 
peak as a function of temperature close to 
the peak in the dielectric constant while the intensity of 
the central peak scattering increases strongly below this temperature. 
These results are discussed in terms of a random field model for relaxors.
\end{abstract}


\submitto{\JPCM}

\maketitle

\section{Introduction}
\label{intro}
\noindent The family of complex perovskites AB$'_x$B$''_{1-x}$O$_3$ 
has been known for fifty years and are 
related to the classic ABO$_3$ perovskites but with different physical 
properties due to the chemical disorder on the B-sublattice. A very 
interesting sub-class of these complex perovskites is formed by the 
relaxor ferroelectrics, or shortly - relaxors. 
These materials have a frequency-dependent peak in the dielectric 
permittivity $\varepsilon'$ which typically extends over hundreds of 
degrees and is not directly related to any macroscopic changes of the 
symmetry. Since many physical properties of relaxors exhibit anomalies 
in this temperature range, the properties were called a "diffuse phase 
transition". Despite many investigations, however, the nature of the 
diffuse phase transition remains 
unclear~\cite{smol1,cross,kleemann, viehland, blinc}. 

PbMg$_{1/3}$Nb$_{2/3}$O$_3$ (PMN) is a model relaxor ferroelectric. It has 
the cubic  Pm$\bar{3}$m structure at all temperatures in the absence of an 
external electric field~\cite{husson1}. However, in an applied electric 
field, PMN undergoes a structural phase transition at T$\sim210$~K~\cite{krainik} 
while the anomaly in the dielectric permittivity $\varepsilon'$ appears 
around the mean Curie temperature T$_{cm}\sim$~270~K~\cite{smol1}.
At a higher temperature, T$_d\sim 620$~K, the optical refractive index 
departs from the expected linear temperature dependence as observed 
by Burns and Scott~\cite{burns}. This result was explained by the 
appearance of small polar regions of size of a few unit cells, which 
were referred to as 'polar nanoregions' (PNR). 

The lattice dynamics of PMN and similar relaxor crystals 
PbMg$_{1/3}$Ta$_{2/3}$O$_3$ (PMT), PbZn$_{1/3}$Nb$_{2/3}$O$_3$ (PZN) and 
PbZn$_{1/3}$Nb$_{2/3}$O$_3$-xPbTiO$_3$ (PZN-xPT) 
has been studied by light~\cite{burns,serg1989,serg1992,siny1,siny2,toulose1} 
and neutron scattering~\cite{sb1,shirane20001,shirane20002,shirane20011,
shirane20012,shirane20021,shirane20023,
shirane20025,sb2,seva20032,kulda2,
seva20033,seva20034,seva20041,seva20042,shirane20042}. 
First-order Raman light scattering was observed in both PMN and PZN~\cite{burns}, 
although first order scattering is forbidden if the crystal structure remains that 
of the cubic perovskites. Until now there is no detailed understanding for the 
origin of the apparent first order Raman scattering in the 
relaxors~\cite{siny2,toulose1}. The search for a soft mode by light 
scattering in PMN was unsuccessful~\cite{burns,siny2}, whereas the results of 
neutron scattering measurements have lead to the suggestion that the soft TO 
phonons are overdamped for $q\lesssim0.12$~rlu and 
$200 {\rm K} \lesssim T \lesssim T_d$~\cite{shirane20002,shirane20021,shirane20025}. 
This anomalous behavior of the TO phonons was attributed to 
the formation of PNR in both PZN and 
PMN~\cite{shirane20011,shirane20012,shirane20021,shirane20025}, 
although an alternative interpretation was recently proposed that describes the 
results and the so called "waterfall" effect in terms of coupling between TA 
and highly damped TO phonons~\cite{kulda2}. In addition, both 
light and neutron spectroscopy have reported a quasi-elastic (QE) component 
in the phonon spectrum of 
PMN~\cite{serg1989,siny1,toulose1,seva20041,seva20042,shirane20042}. 
As anharmonicity is important in relaxors, it is {\it a priori} possible that 
this quasi-elastic mode originates from the strong coupling between the TA 
and heavily damped TO phonon modes~\cite{cowley19801}. 

In this work we have used high resolution neutron scattering to investigate the 
low-energy part of the vibration spectrum of PMN as described in section 2. 
We find in section 3 that the lowest transverse optic (TO) phonon branch is 
strongly coupled to the transverse acoustic branch (TA) above the 
Burns temperature, T$_d$, where the QE component is absent. Below T$_d$, the 
results suggest that the phonon spectrum remains essentially unchanged, with 
the TO phonons found to be under-damped throughout the temperature range 
and the "waterfall effect" is not clearly observed at least in the [0,0,q] 
direction. In addition there are two more components of the scattering 
consisting of a quasi-elastic, QE, component whose energy width is larger than 
the resolution and a central peak, CP, component whose energy width is the 
same as the resolution function. The energy and wave-vector width of the 
QE component have been measured and it is found that the QE susceptibility is 
a maximum at a temperature of about 370 K, a somewhat larger temperature 
than the maximum in the dielectric susceptibility. The CP component increases 
rapidly in intensity below 370 K similarly to the behaviour expected from 
an order parameter. The line-shape of the scattering as a function of wave-vector 
is not however a sharp Bragg reflection but is approximately a Lorentzian 
to the power of 1.5. In section 4 we shall suggest that these results are 
consistent with a random field model for this relaxor.
\section{Experimental}
\label{experiment}
\noindent The measurements were carried out with the three-axis spectrometer 
TASP, located at the neutron spallation source SINQ at the Paul Scherrer 
Institut, Switzerland. A high-quality single crystal of PMN ($\sim$ 8~cm$^3$) 
was mounted using a niobium holder inside a furnace. The measurements were 
performed in the temperature range 150~K - 670~K. 
The crystal was aligned in the (h~h~l)-scattering plane. 
The spectrometer had a pyrolytic graphite (PG) monochromator and analyzer 
and was operated in the constant final-energy mode. In the course of 
the measurements we had to search for compromises between the spectrometer 
resolution, the accessible ($Q-\omega$) range, and the intensity. As a result 
several different spectrometer configurations were used. Most of data were 
collected in the following three configurations:\\ 
\begin{itemize}
\item (i) $k_{f}=1.64$ \rm \AA$^{-1}$ and $10'/\rm \AA-80'-80'-80'$ collimation
\item (ii) $k_{f}=2.662$ \rm \AA$^{-1}$ and $10'/\rm \AA-80'-80'-80'$ collimation
\item (iii) $k_{f}=2.662$ \rm \AA$^{-1}$ and $10'/\rm \AA-20'-20'-80'$ collimation
\end{itemize}
where $k_{f}$ refers to the energy of the scattered neutrons.
In configuration (i) the (0~0~2) Bragg reflection of PG was used for both the 
monochromator and analyzer, whereas the (0~0~4) reflection was used for 
configurations (ii) and (iii). The energy resolution at zero energy transfer 
was 0.22 meV for configuration (i), 0.5 meV for (ii), and 0.25 meV for (iii), 
respectively. For configuration (i) the q-resolution along the scattering 
vector was q$_{par}$=0.015~$\rm \AA^{-1}$ and q$_{perp}$=0.021~$\rm \AA^{-1}$ 
in the transverse direction. For configurations (ii) and (iii) 
q$_{par}$=0.02~$\rm \AA^{-1}$ and q$_{perp}$=0.04~$\rm \AA^{-1}$, respectively. 
The vertical resolution was q$_z$=0.1 $\rm \AA^{-1}$ with $k_f = 1.64 \rm \AA^{-1}$ 
and q$_z = 0.16 \rm \AA^{-1}$ with $k_f = 2.662 \rm \AA^{-1}$, respectively. 
\section{Experimental results and data analysis}
\label{treat}
\subsection{Measurements above the Burns temperature at $T=670$~K}
\label{dispersions670}
\begin{figure}[h]
\centering
\includegraphics[width=0.45\textwidth]{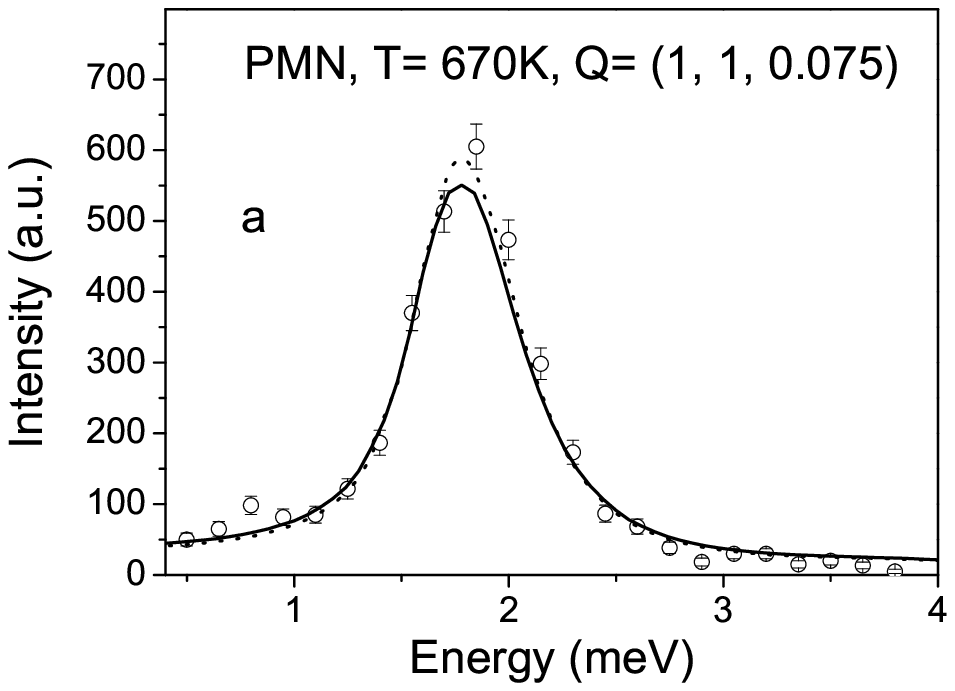}
\includegraphics[width=0.45\textwidth]{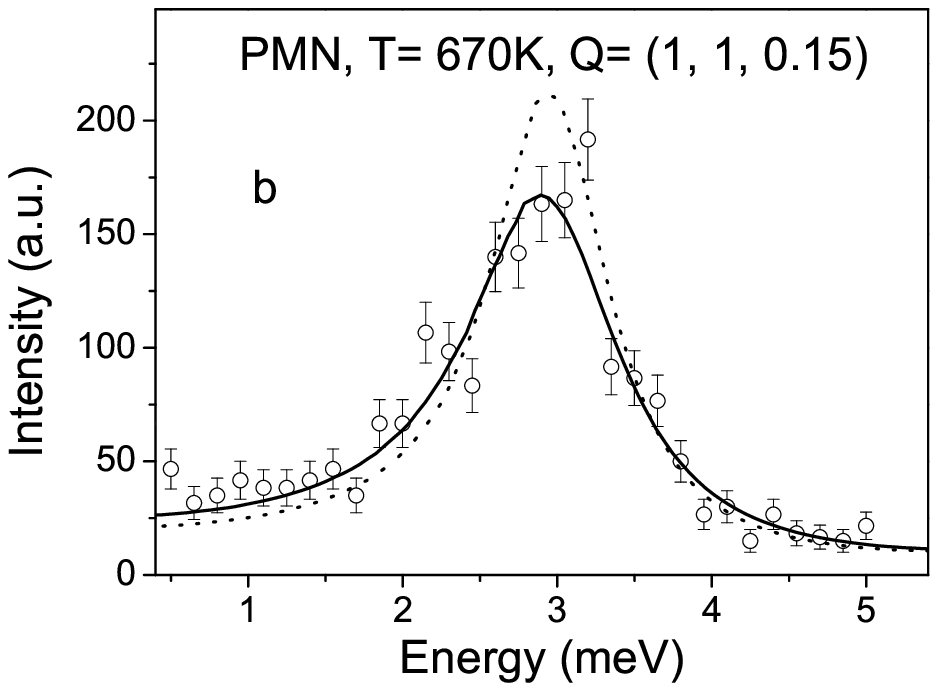}
\includegraphics[width=0.45\textwidth]{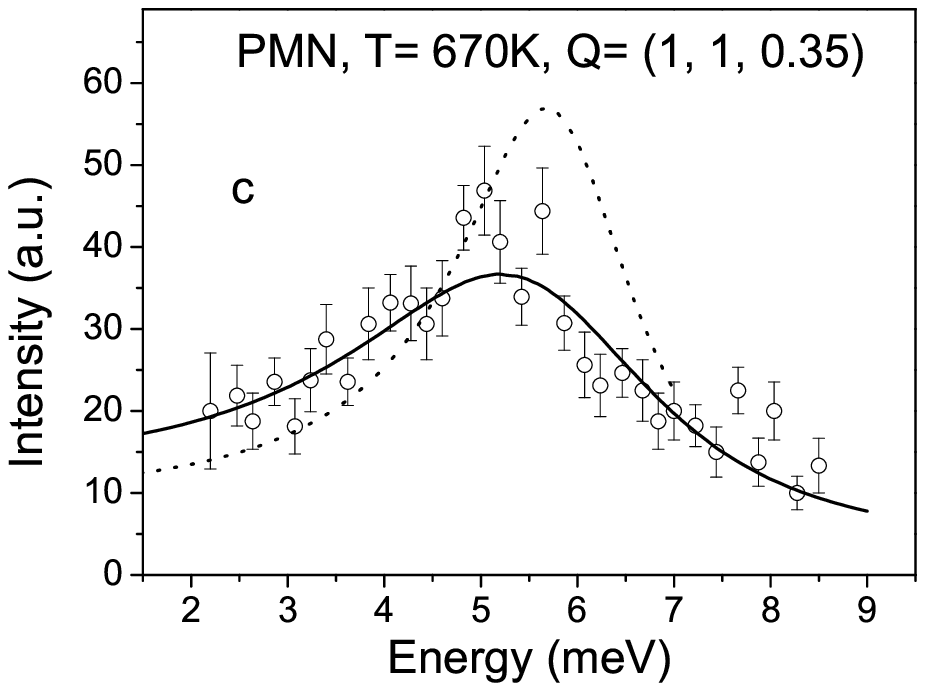}
\includegraphics[width=0.45\textwidth]{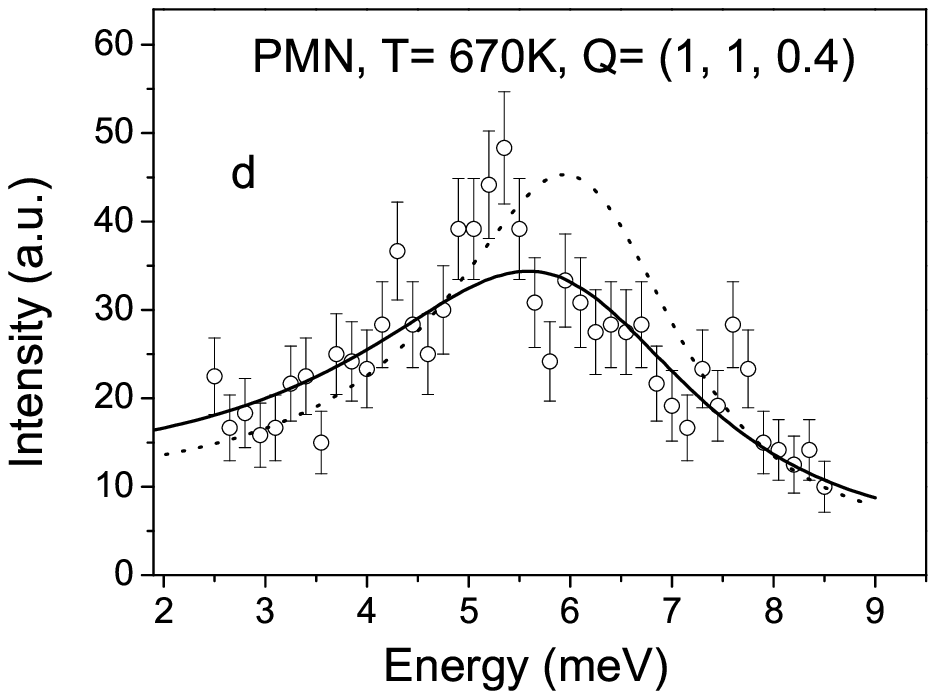}
\caption{The results of typical constant-Q scans using configuration 
          (i) in the (1,~1,~0) BZ. The solid line shows 
          the fit results whereas the dotted line 
          corresponds to simulated spectra with the 
          imaginary part of the coupling set to zero. 
          See text for details. $q$ is given in  
          reciprocal lattice units (rlu). 
          1 rlu=1.55 $\rm\AA^{-1}$.   
         }
\label{fig1}
\end{figure}
%
\begin{figure}[h]
\centering
  \includegraphics[width=0.45\textwidth]{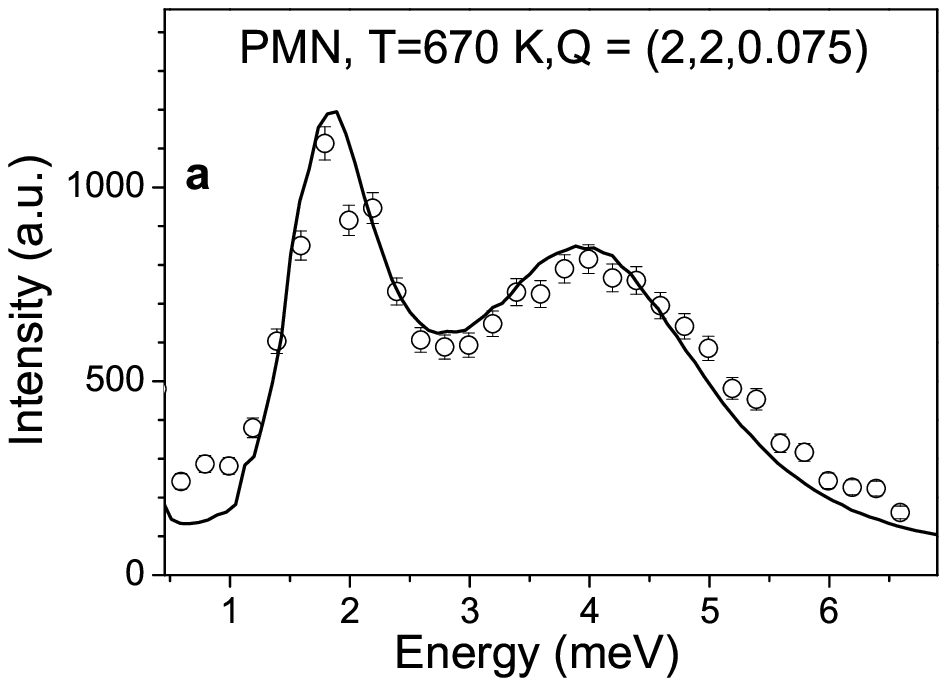}
  \includegraphics[width=0.45\textwidth]{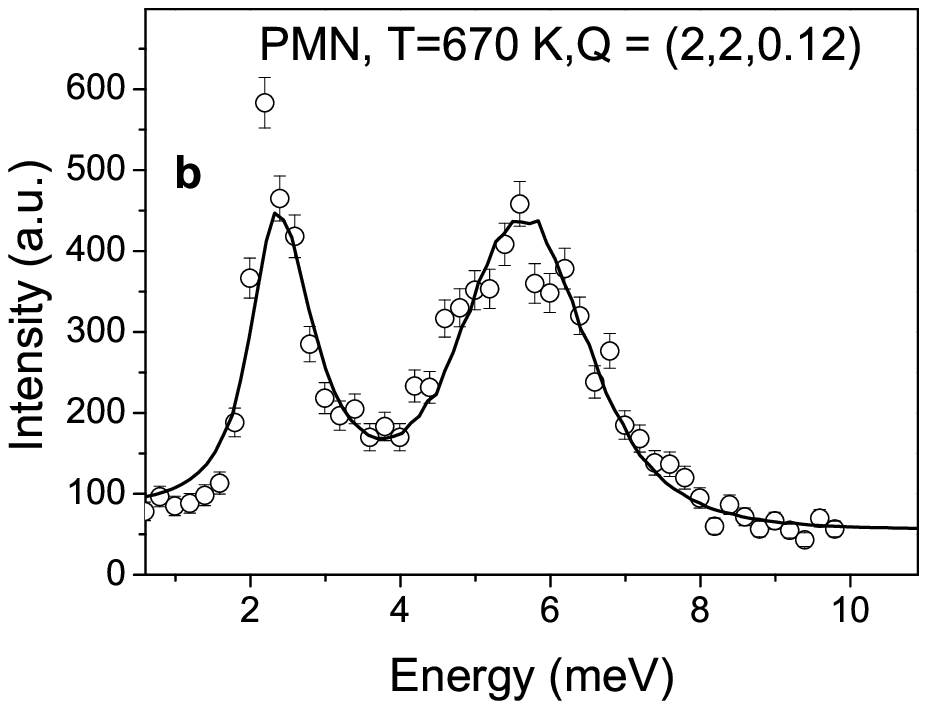}
  \includegraphics[width=0.45\textwidth]{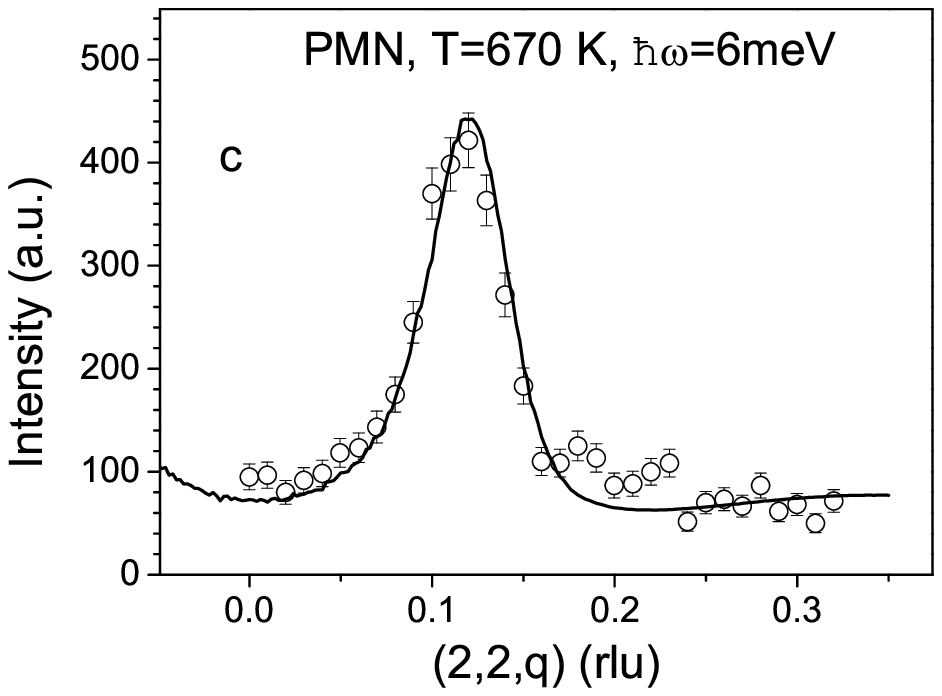}
  \includegraphics[width=0.45\textwidth]{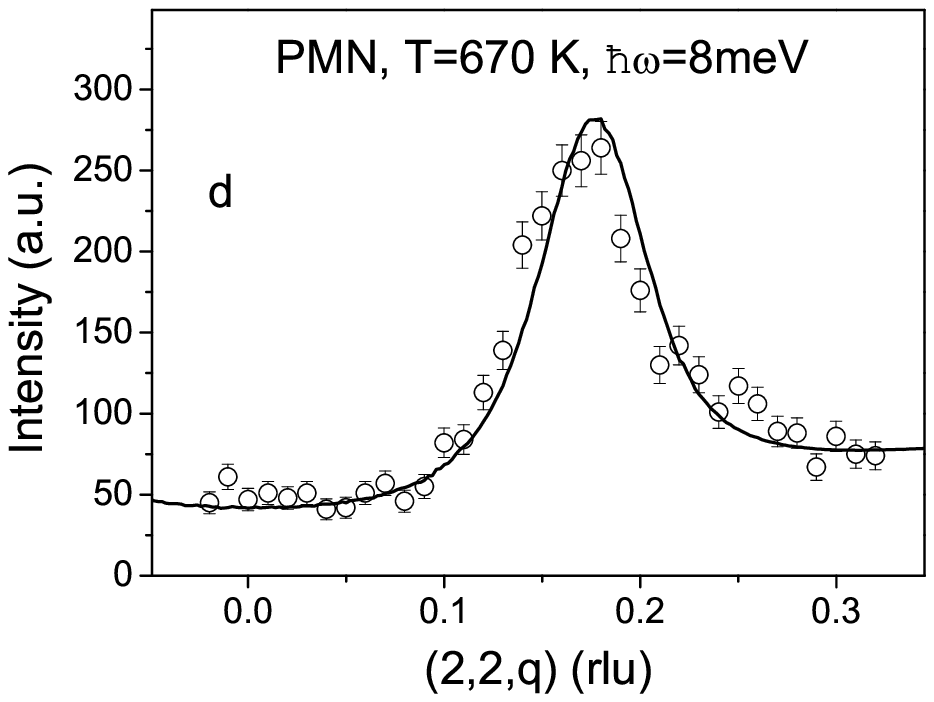}
  \caption{The results of typical constant-Q and 
           with the results of constant-E   
           scans taken in the (2,~2,~0) BZ using  
           configuration (ii) with results of fits 
           as explained in the text. 
           }
\label{fig2}
\end{figure}
\begin{figure}[h]
\centering
  \includegraphics[width=0.6\textwidth]{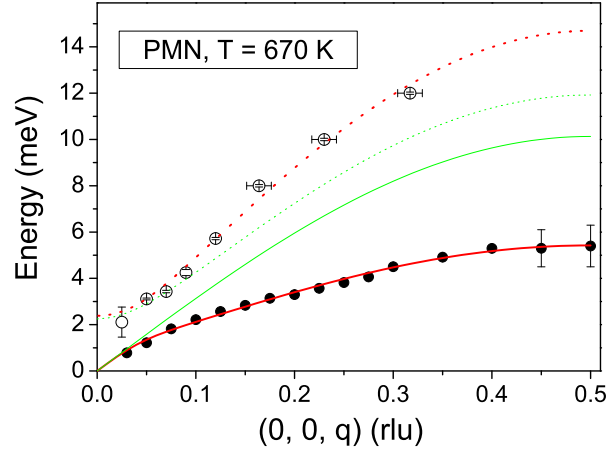}
  \caption{(Color online) Observed and calculated dispersion 
           of the TA (solid circles) and the lowest TO 
           (open circles) phonons in PMN along the [00q] 
           direction measured at $T=670$~K. 
           The solid and dotted green lines correspond to 
           the calculated dispersion of the TA and TO 
           phonons with coupling set to zero. 
           }
\label{fig3}
\end{figure}
%
\begin{figure}[h]
\centering
  \includegraphics[width=0.6\textwidth, angle=0]{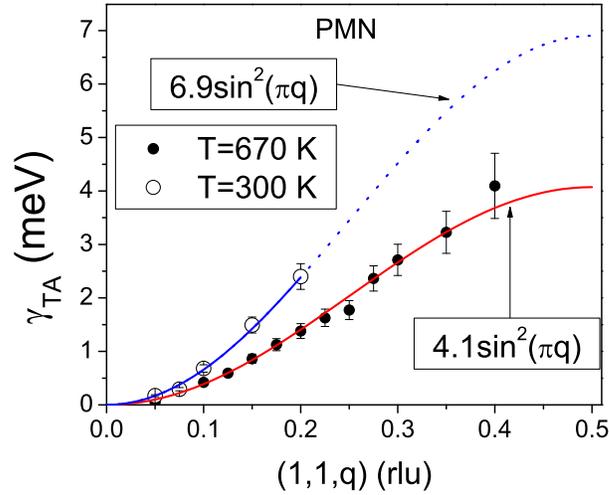}
  \caption{(Color online) $q$-dependence of the damping of the transverse 
           acoustic phonon $T=670$~K and $T=300$~K 
           (see text for details).
           }
\label{fig4}
\end{figure}
%
%
\noindent The dynamical susceptibility for two coupled excitations 
was considered, for example in Refs.~\cite{cowley19731,cowley19801,wehner} 
and is given by 
\begin{equation}
\label{ccm}
\chi_{CM}({\bf Q},\omega)=\frac{f_1^2\chi_1+f_2^2\chi_2+2\lambda f_1f_2\chi_1\chi_2}
{1-\lambda^2\chi_1\chi_2}, 
\end{equation}
where $\chi_i\equiv\chi_i({\bf Q},\omega)$, $i=1,2$ 
are the dynamical susceptibilities of the uncoupled phonons; and the $f_i$ are the 
wave-vector dependent structure factors for the modes. The interaction term contains 
both real and imaginary parts and is assumed to have the simplified form: 
$\lambda\equiv\Delta+i\Gamma_{12}=(\Delta_{12}+i\omega{\gamma_{12}})
\sin^2(\pi q)$ where the q-dependence is chosen to be proportional to q$^2$ at 
small q as expected for a centrosymmetric crystal~\cite{cowley19731}. 
The dynamical susceptibilities of the TA and TO phonons are given by the 
damped-harmonic-oscillator functions (DHO):
\begin{equation}
\label{dho}
\chi_{DHO}({\bf q},\omega)=(\omega^2_q-i\gamma_q\omega-\omega^2)^{-1}.
\end{equation} 
In Eq.~\ref{dho}, $\gamma_q$ is the damping and $\omega_{q}$ is 
the renormalized frequency of a phonon. 

When the damping is neglected the resonance frequency of the two 
coupled modes is given by: 
 \begin{equation}
 \label{anticross}
 \Omega_{1,2}^2=\frac{1}{2}\left( \omega_1^2+\omega_2^2\pm
 \sqrt{(\omega_1^2-\omega_2^2)^2+4\Delta}\right),
 \end{equation} 
where $\omega_1$ and $\omega_2$ are the frequencies of uncoupled excitations. 
To model the dispersion of the TA and TO phonons we have used the following 
parameterization based on a low q expansion 
 \begin{equation}
 \label{dispta}
 \omega_1\equiv\omega_{TA}=d\sin(\pi q), 
 \end{equation} 
and
 \begin{equation}
 \label{dispto}
\omega_2^2\equiv\omega_{TO}^2=\omega_{TO}^2(0)+c\sin^2(\pi q). 
 \end{equation} 
\noindent The damping of TA phonon was found to increase with q approximately as 
\begin{equation}
\label{gta}
\gamma_{TA}=D_{TA}\sin^2(\pi q).
\end{equation}
and no assumption was made about the damping of the TO phonon. 

The neutron scattering intensity $I(\mathbf{Q},\omega)$ was modelled as:  
\begin{equation}
\label{intensity}
I(\mathbf{Q},\omega)=S(\mathbf{Q},\omega){\otimes}\,R(\mathbf{Q},\omega)+B.
\end{equation}
\noindent The symbol $\otimes$ stands for the 4-D convolution with the 
spectrometer resolution function $R(\mathbf{Q},\omega)$~\cite{popa}, 
while $B$ denotes the background level and $S(\mathbf{Q},\omega)$ 
is the neutron scattering function which is related to the imaginary part 
of the dynamical susceptibility, $\chi''(\mathbf{Q},\omega)$, through  
\begin{equation}
\label{sqw2}
S(\mathbf{Q},\omega)={[n(\omega)+1]\over{\pi}}\chi''_{CM}(\mathbf{Q},\omega) 
\end{equation}
with the temperature factor $[n(\omega)+1]=[1-\exp(-\hbar\omega/k_BT)]^{-1}$. 

\begin{figure}[h]
\centering
\includegraphics[width=0.60\textwidth]{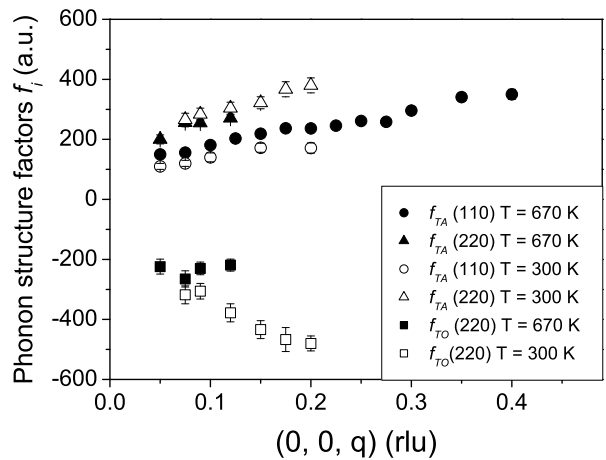}
\caption{Wavevector dependence of the structure factors of 
         the TA and TO phonons  
         along the [0,~0,~q] direction. 
        }
\label{fig5}
\end{figure}

Figures~\ref{fig1} and \ref{fig2} show typical inelastic neutron 
scattering spectra of PMN at $T=670$~K in the (1,~1,~0) and (2,~2,~0) 
Brillouin zones (BZ) and the results of fitting these spectra to the 
interacting phonon model, Eq.~\ref{sqw2}. In the (1,~1,~0) BZ only the TA phonon 
is observed in constant Q scans, whereas both TA and TO phonons are visible 
in the (2,~2,~0) BZ. The values of parameters which correspond to the solid lines in 
Figs.~\ref{fig1} and~\ref{fig2} are: 
$d=10.05\pm0.45$~meV, $D_{TA}=4.1\pm0.5$~meV, $\omega_{TO}(0)=2.1\pm0.2$~meV, 
$c=137\pm9$~meV$^2$, $\Delta_{12}=85\pm7$~meV$^2$, 
$\gamma_{12}=1.5\pm0.5$~meV. The wavevector dependence of the damping of the TO 
phonon, $\gamma_{TO}$ is given in Table~\ref{tab1}. The TA and TO dispersion curves 
are correctly reproduced with this parameterization and the frequencies are shown 
in Fig.~\ref{fig3}. Also shown in Fig.~\ref{fig3} is the dispersion of the TA and 
TO branches in the absence of coupling. The two branches do not cross and the 
dispersion of both branches is mostly affected by the coupling close to the zone 
boundary. As illustrated in Fig.~\ref{fig1} the imaginary part of the coupling 
affects the phonon line-shape mainly at large momentum transfers. 
Figure~\ref{fig4} shows that the damping of the TA phonon is proportional to 
$\sin^2(\pi q)$ that was obtained from fitting the spectra individually. 
Finally, Fig.~\ref{fig5} shows the structure factor of the TA and TO modes. 
%
%
%
\begin{table}[t]
\centering
\caption{Wave-vector dependence of $\gamma_{TO}$ in PMN.}
\begin{tabular}{|c|ccccccc|} 
\hline 
q (rlu)           & 0.05  & 0.075  & 0.09  & 0.12 & 0.15 & 0.175  & 0.2\\
\hline 
$\gamma_{TO}(670 K)$(meV) &0.8   & 1.4    & 1.8   & 1.3  &  -   &  -     &  -  \\
\hline
Std. Error         &0.1   &0.2     &0.3    & 0.2  & -    &  -     &  - \\
\hline                                                                                                
$\gamma_{TO}(300 K)$(meV)& 1.0   & 1.8    & 2.0   & 2.4  & 3.9  & 4.2    & 4.4\\
\hline                                                                                                           
Std. Error        & 0.2   & 0.3    & 0.4   & 0.4  & 0.7  & 0.9    & 0.9\\
\hline                                                                                                           
\end{tabular}
\label{tab1}
\end{table} 
%
\subsection{Temperature dependence of the vibration spectrum} 
\label{tdata}
%
\begin{figure}[h]
\centering
  \includegraphics[width=0.45\textwidth]{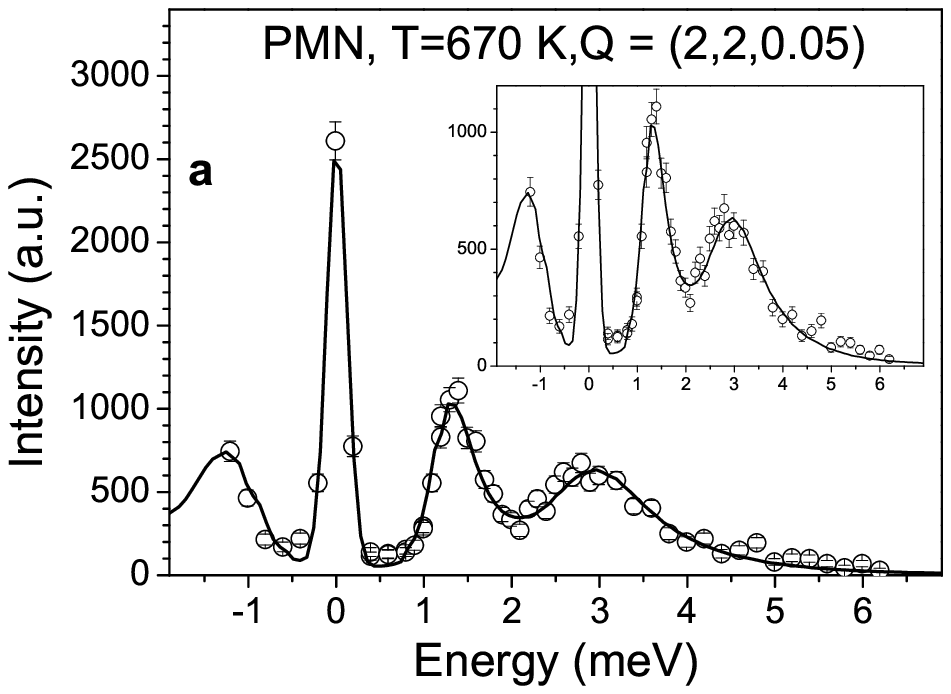}
  \includegraphics[width=0.45\textwidth]{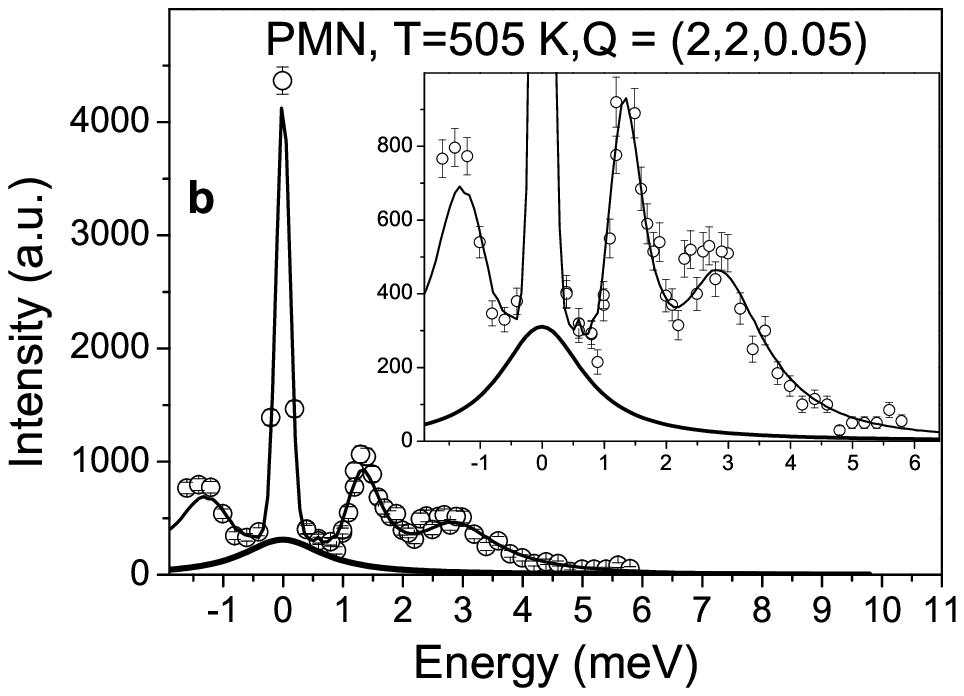}
  \includegraphics[width=0.45\textwidth]{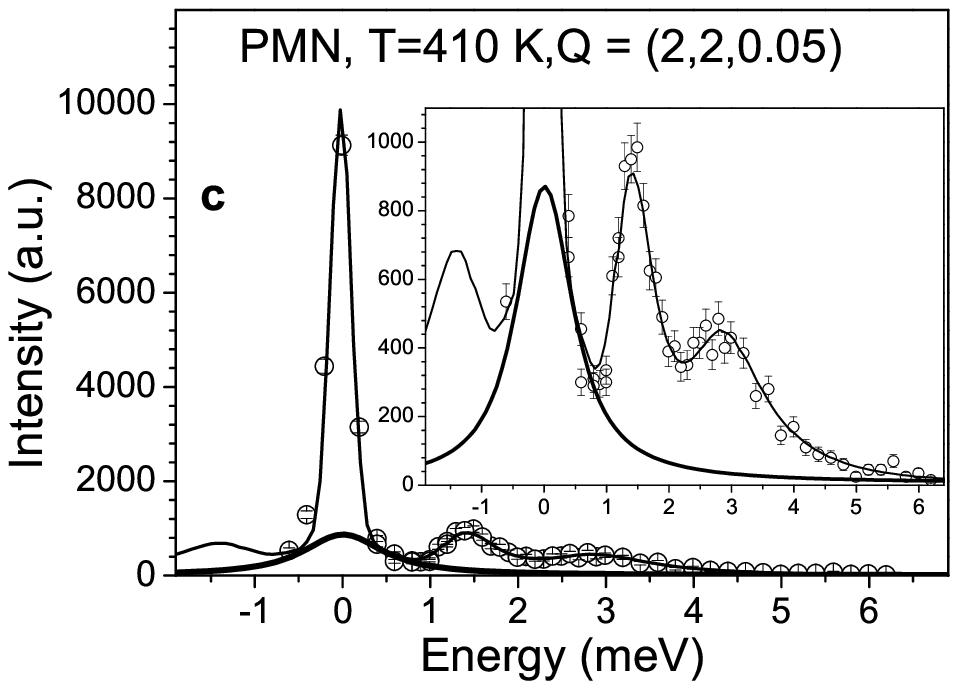}
  \includegraphics[width=0.45\textwidth]{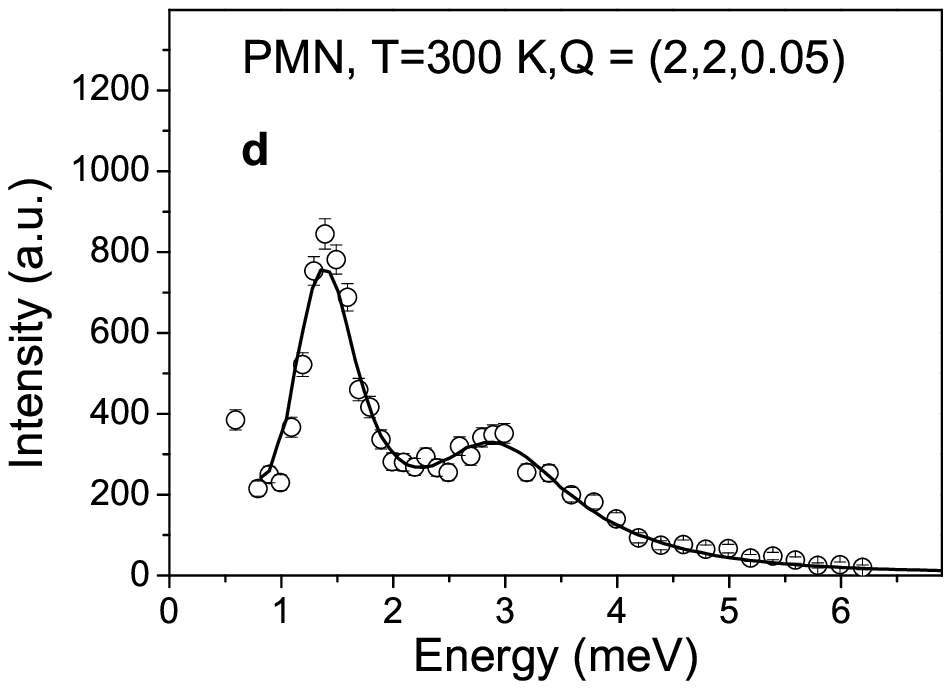}
  \caption{Observed and calculated spectra from 
           constant-$Q$ scans taken 
           at $\mathbf{Q}$=(2,~2,~0.05) as a function of temperature.
           The contribution of QE scattering is shown by 
           a bold solid line. 
           }
\label{fig6}
\end{figure}
%
\begin{figure}[h]
\centering
  \includegraphics[width=0.45\textwidth]{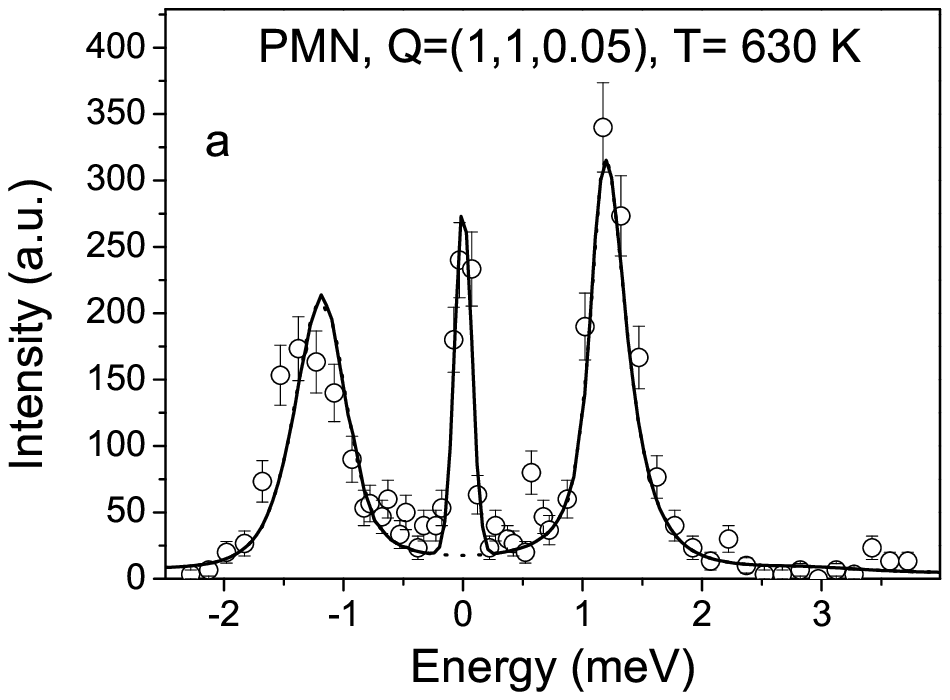}
  \includegraphics[width=0.45\textwidth]{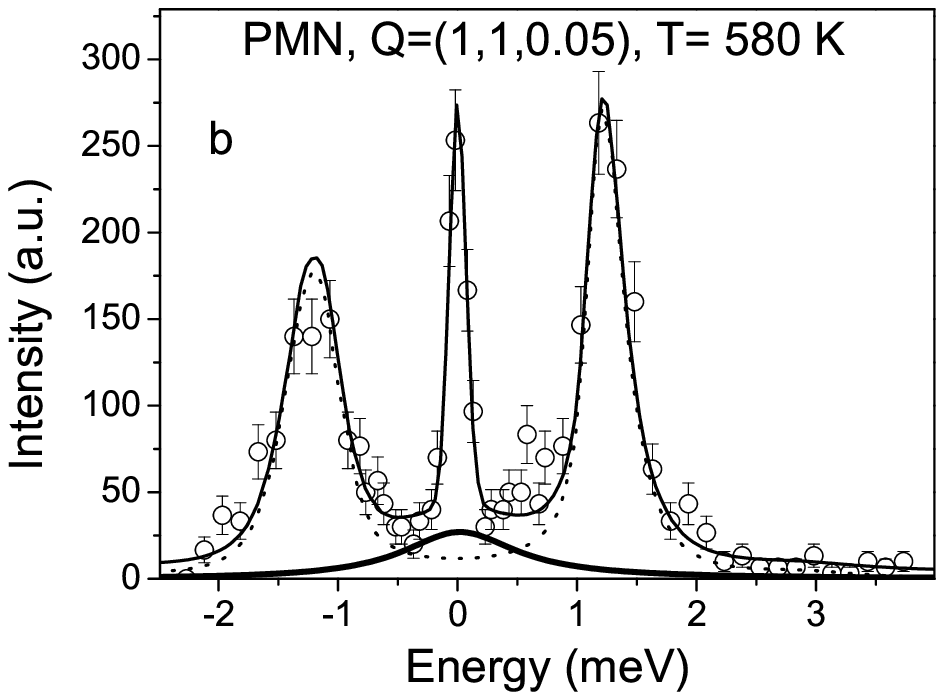}
  \includegraphics[width=0.45\textwidth]{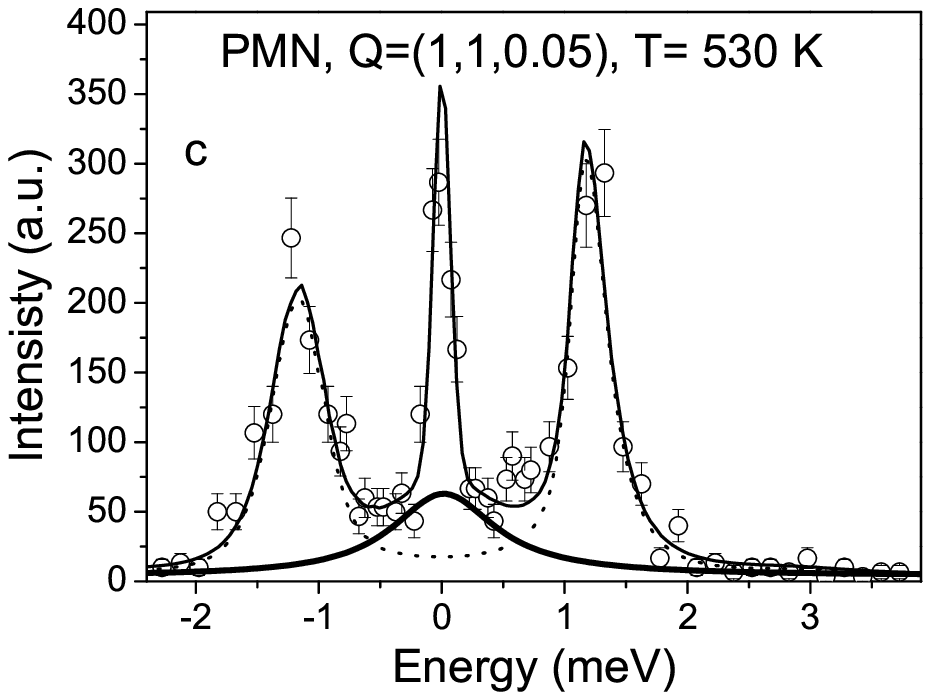}
  \includegraphics[width=0.45\textwidth]{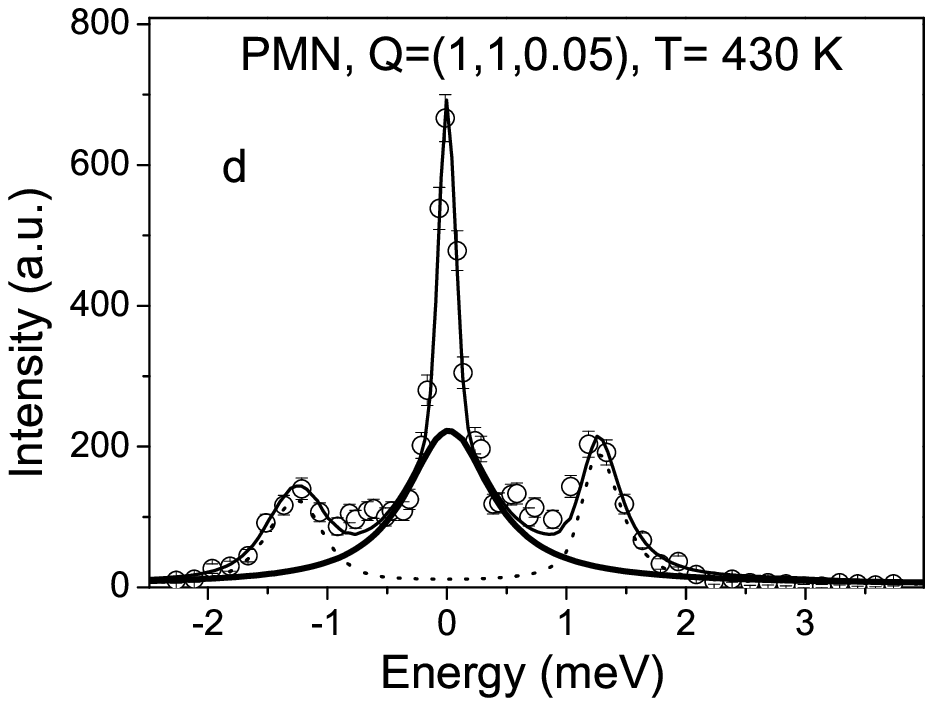}
  \includegraphics[width=0.45\textwidth]{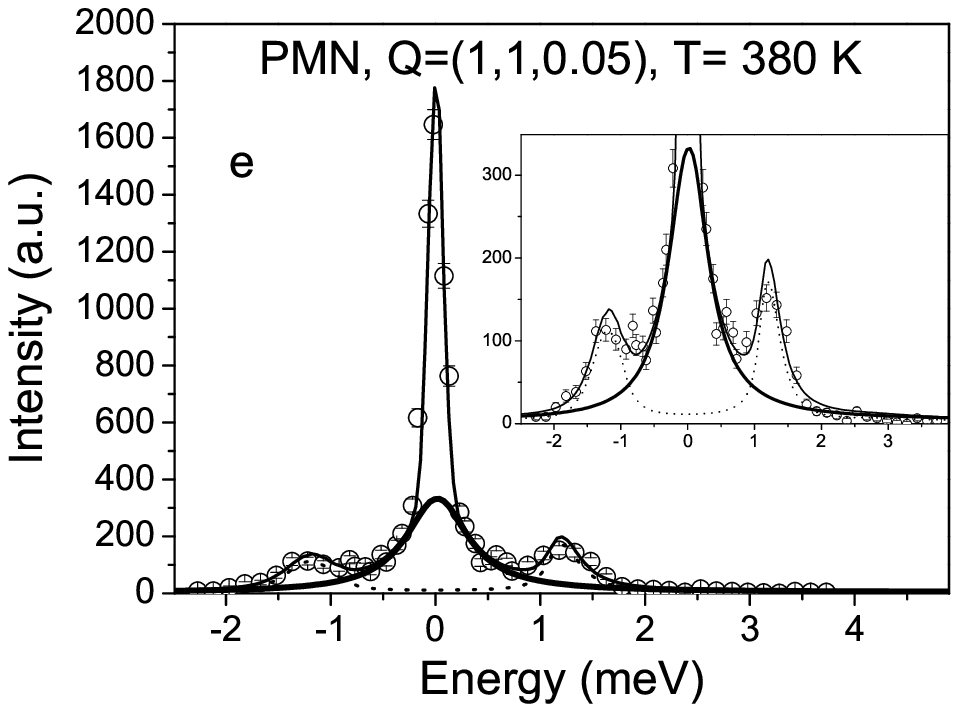}
  \includegraphics[width=0.45\textwidth]{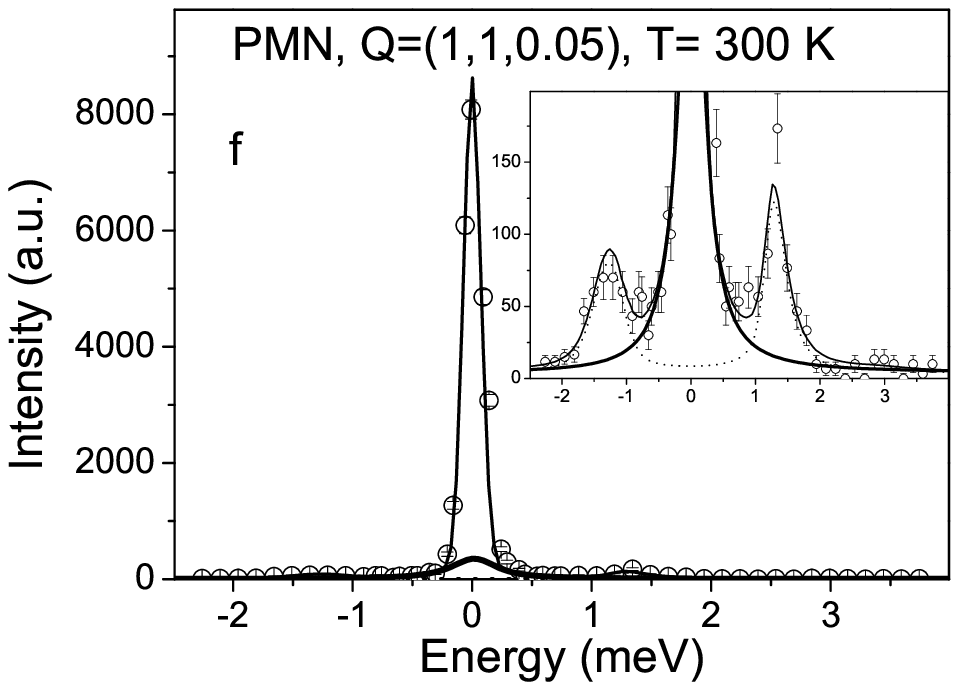}
  \caption{Observed and calculated spectra from constant-$Q$ scans taken 
           at $\mathbf{Q}$=(1,~1,~0.05). The contribution 
           of QE scattering is shown by a bold solid line. 
           Note the increase of CP intensity as the temperature 
           is lowered below $T_d$.
           These measurements were performed using $k_{f}=1.64$ 
           \AA$^{-1}$ and $10'/\rm \AA-20'-20'-80'$ collimation.
           }
\label{fig7}
\end{figure}
%
\noindent The energy and $q$ space resolution was improved to study the 
temperature dependence of the TA and TO modes close to the Brillouin zone 
center. Figures~\ref{fig6} and~\ref{fig7} show 
examples of spectra measured at $\mathbf{Q} = (2,~2,~0.05)$ and 
$\mathbf{Q} = (1,~1,~0.05)$, respectively. At 670 K, apart from the resolution 
limited elastic scattering, only scattering from the TA and TO phonons is 
observed. Upon decreasing the temperature, additional quasi-elastic scattering, 
QE, becomes visible with an energy width larger than the resolution function 
while the more intense strictly elastic scattering, the CP component, 
increases on cooling especially below a temperature of 370 K. 
Both components of this elastic scattering have been observed previously in 
PMN~\cite{seva20041,seva20042} 
and can be described by a resolution limited central peak (CP) and a Debye-like 
relaxation mode. The neutron-scattering susceptibility is given by: 
\begin{eqnarray}
\label{sqwfinal}
S(\mathbf{Q},\omega)&=&S_{CP}({\bf Q},\omega)  \\ 
&+&\frac{[n(\omega)+1]}{\pi}
[\chi_{CM}''({\bf Q},\omega)+f_{QE}^2(\mathbf{Q})\chi_{QE}''({\bf Q},\omega)], \nonumber
\end{eqnarray} 
\begin{equation}
\label{cp}
S_{CP}=A(\mathbf{Q})\delta(\omega),
\end{equation} 
\begin{equation}
\label{qe}
\chi_{QE}({\bf q},\omega)=\frac{\chi(0,T)}{1+q^2\xi^2}\cdot(1-i\omega/\Gamma_q)^{-1},
\end{equation} 
where $\delta(\omega)$ is the Dirac function; $f_{QE}$ is the 
structure factor of the QE component and $\chi(0,T)$ is the 
susceptibility at $q=0$; $\xi$ is the correlation length 
of the QE scattering and $\Gamma_q=\Gamma_0+D_{QE}q^2$. 
$S_{CP}$ describes the resolution-limited central peak with intensity $A(\mathbf{Q})$. 

The model, described by Eqs.~\ref{intensity}-\ref{qe}, was fitted to the neutron 
scattering at various temperatures and it gave good agreement with the 
results provided that the elastic scattering components were included. Typical 
results are shown in Figs.~\ref{fig6} and~\ref{fig7}. In the temperature range 
between 300 K and 670 K for q = 0.05 rlu, the parameters describing the phonon 
line-shape remain essentially unchanged for this small wave-vector. However, 
the damping and the amplitude of the QE scattering change significantly as will 
be discussed in the next sub-section. 

\begin{figure}[h]
\centering
  \includegraphics[width=0.45\textwidth]{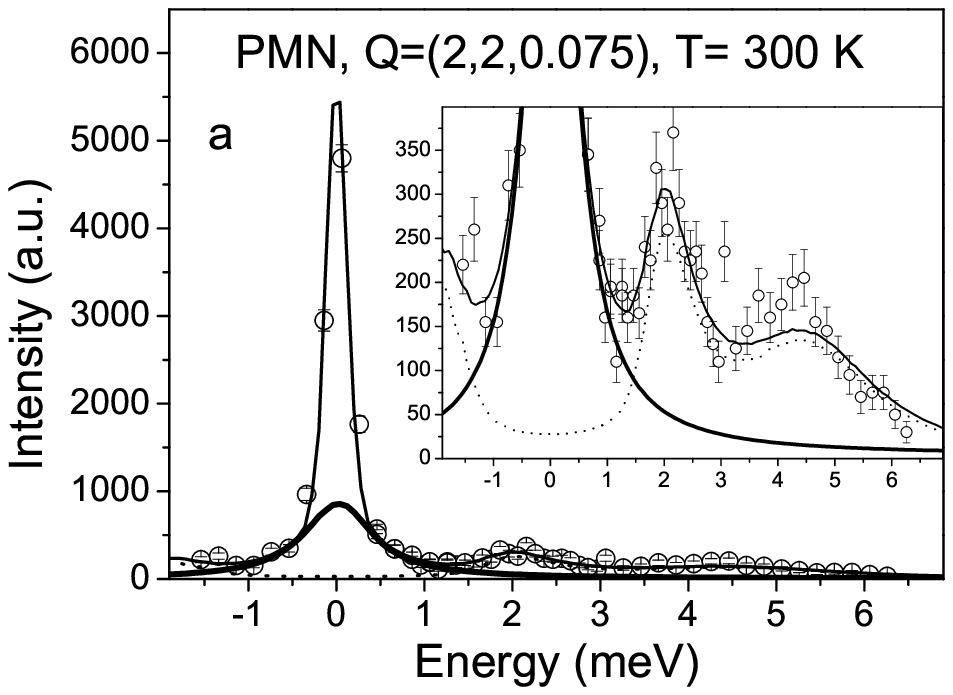}
  \includegraphics[width=0.45\textwidth]{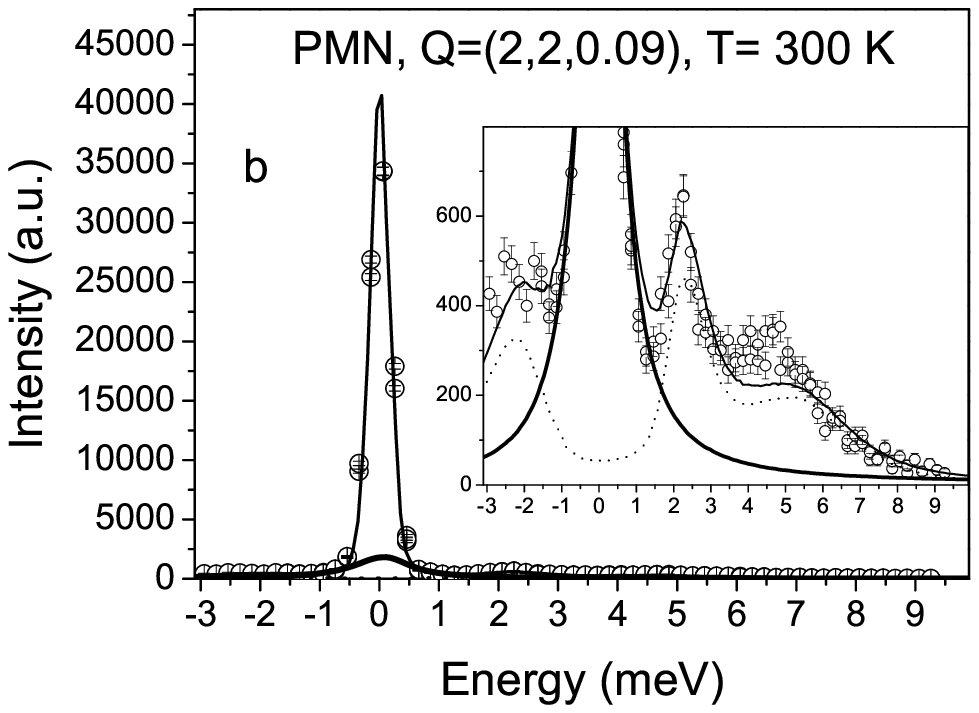}
  \includegraphics[width=0.45\textwidth]{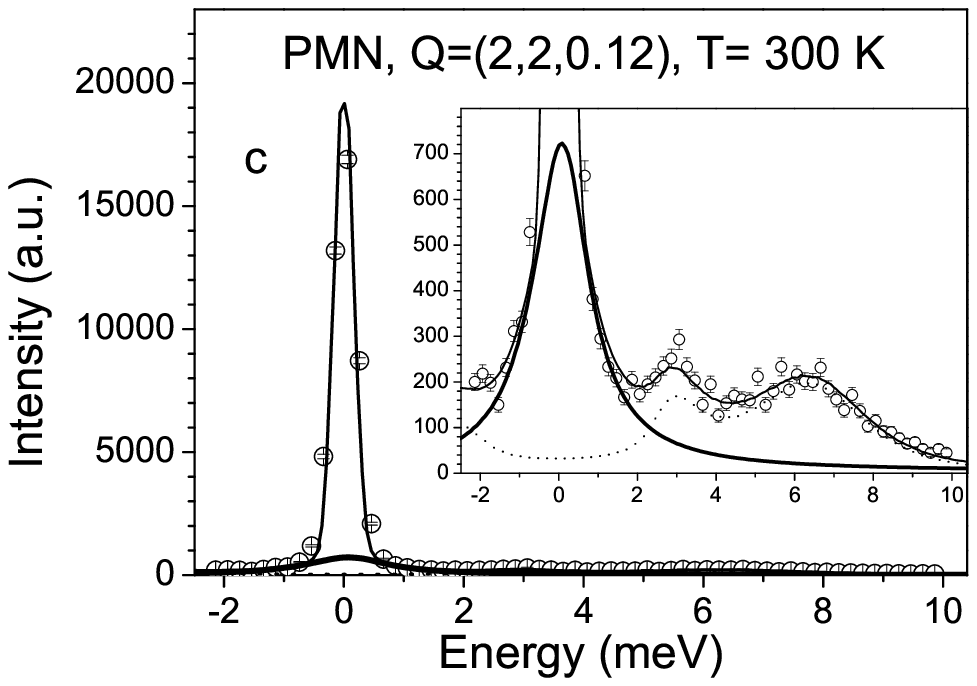}
  \includegraphics[width=0.45\textwidth]{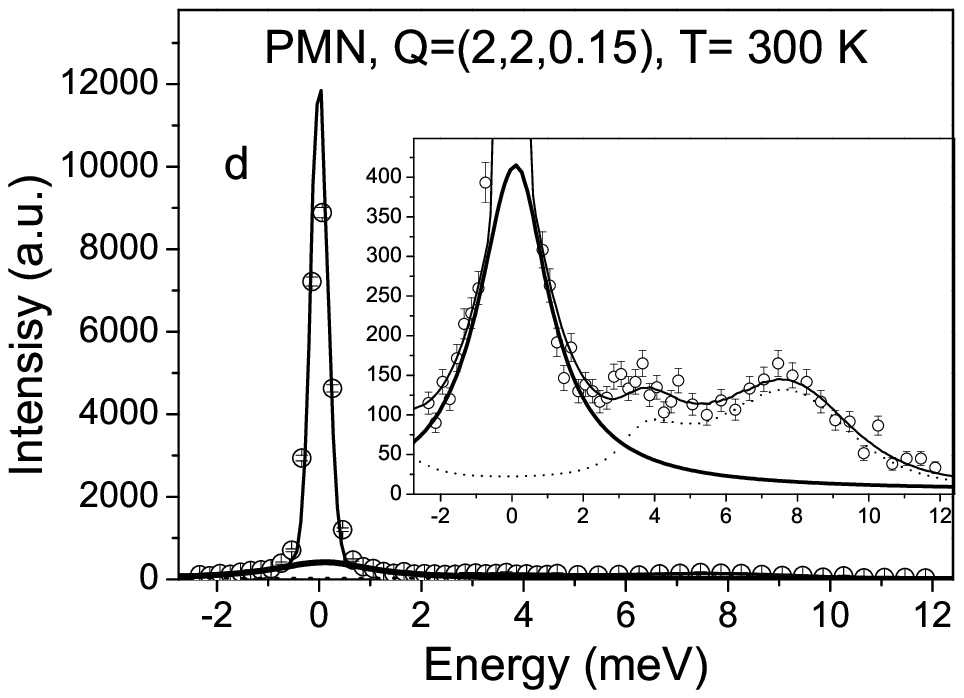}
  \includegraphics[width=0.45\textwidth]{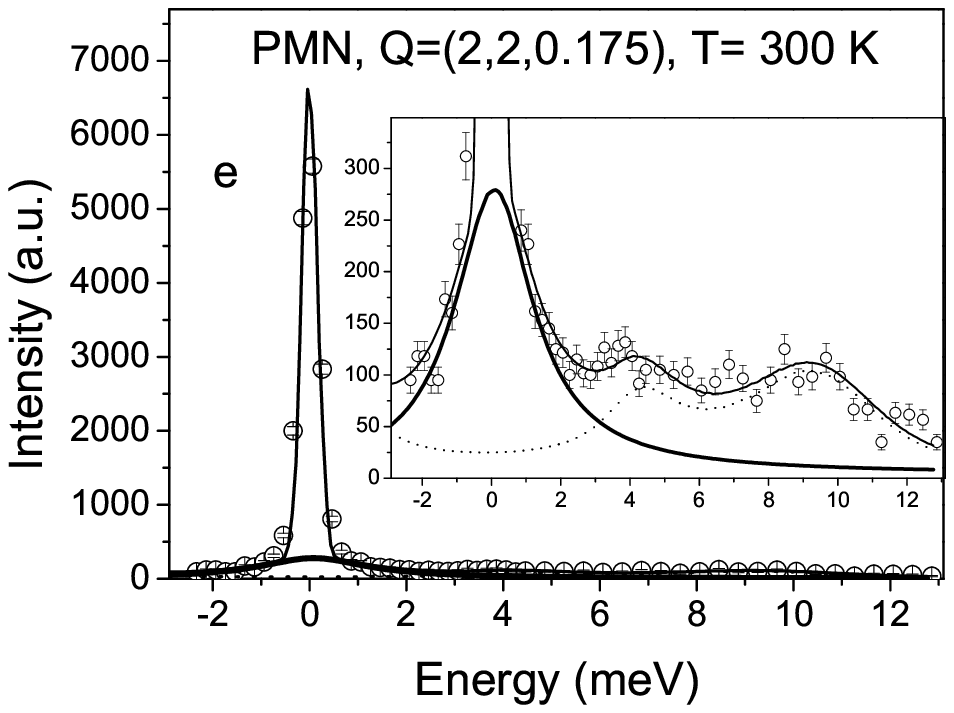}
  \includegraphics[width=0.45\textwidth]{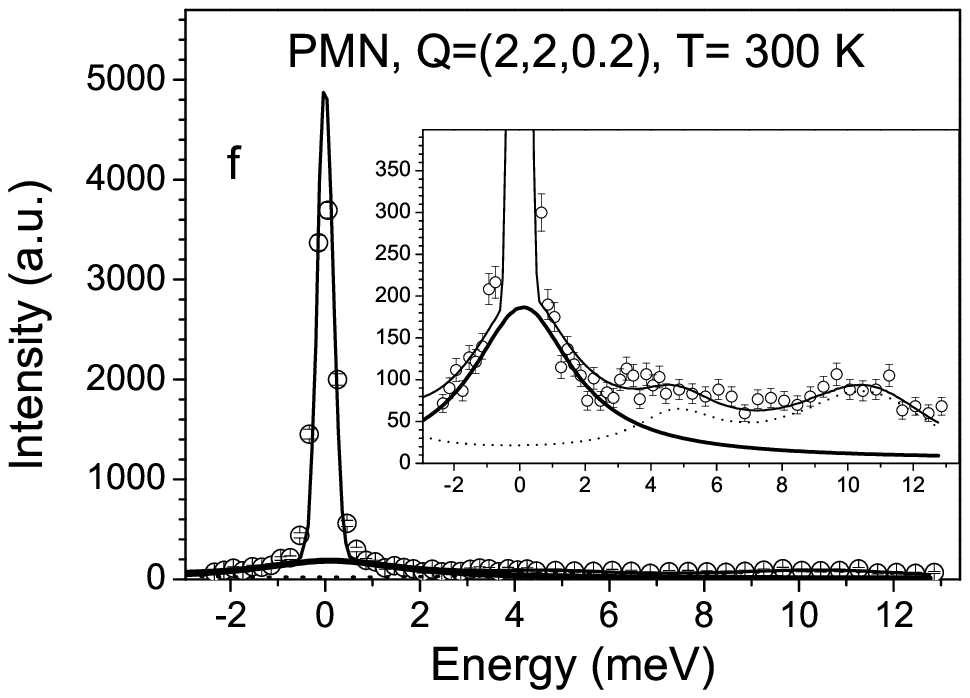}
  \caption{Observed and calculated spectra from constant-$Q$ scans taken 
           in the (2,~2,~0) BZ at T=300 K. Contribution 
           of the QE component is emphasized by bold solid lines. 
           The spectrum shown in Fig.~\ref{fig10}a was measured 
           with configuration (iii); other spectra were measured 
           with configuration (ii).
           }
\label{fig8}
\end{figure}

\begin{figure}[h]
\centering
  \includegraphics[width=0.45\textwidth]{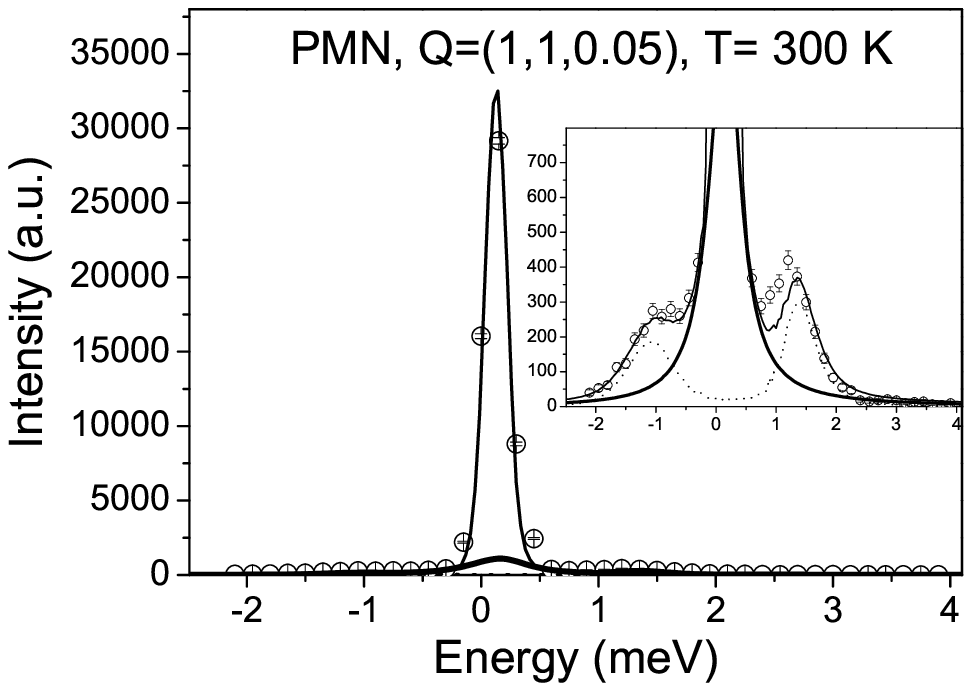}
  \includegraphics[width=0.45\textwidth]{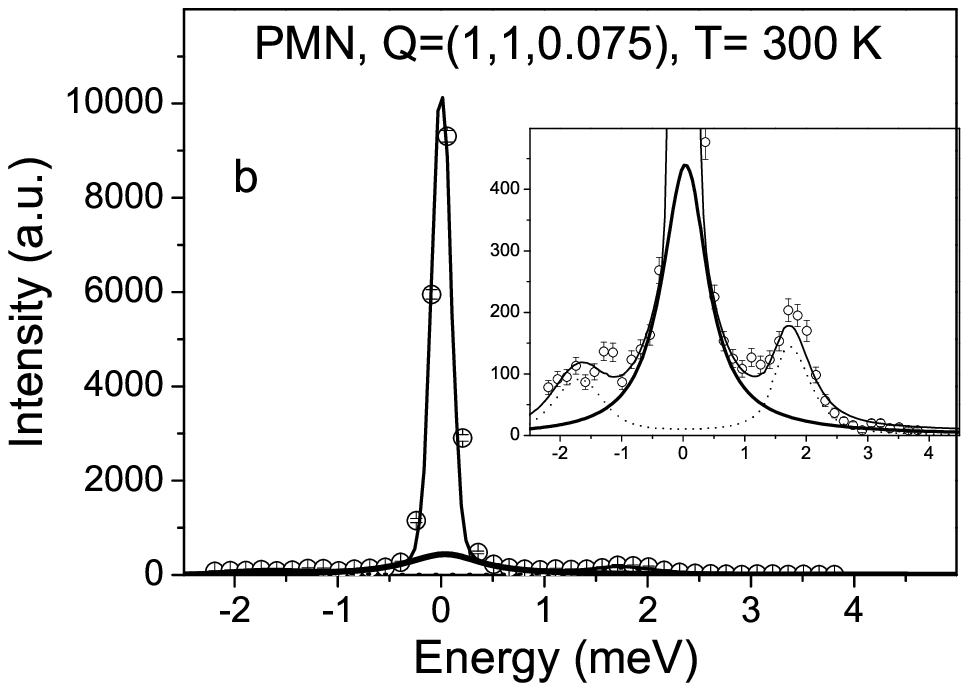}
  \includegraphics[width=0.45\textwidth]{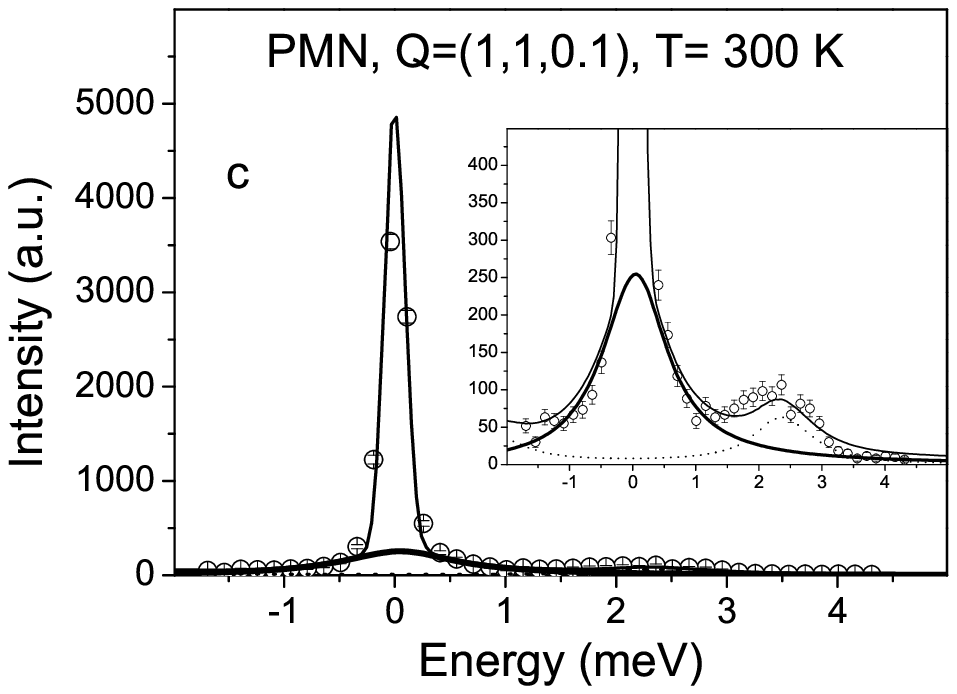}
  \includegraphics[width=0.45\textwidth]{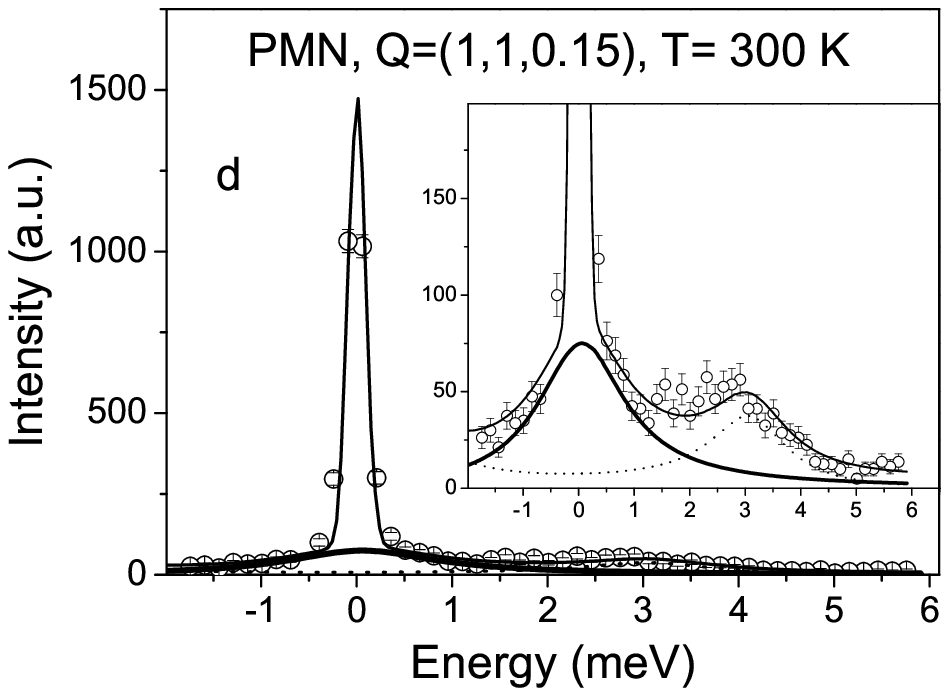}
  \caption{Observed and calculated spectra from constant-$Q$ scans taken 
           in the (1,~1,~0) BZ and configuration (i). Contribution 
           of the QE component is emphasized by bold solid lines.
           }
\label{fig9}
\end{figure}

A more detailed measurement of the scattering as a function of wave-vector was 
performed with constant-Q scans around both the (1,~1,~0) and (2,~2,~0) BZ at 
$T = 300$~K. The results for the scattering and the fits to the model are shown 
in Figs.~\ref{fig8} and~\ref{fig9}. We observe an increase of the damping of the TA 
and TO phonons and of the stiffness of the TO branch. The values of TA damping and 
TO stiffness at $T = 300$~K are $D_{TA} = 6.9\pm0.6$~meV and $c = 260\pm25$~meV$^2$; 
the momentum dependence of the TO-phonon damping is summarized in Table~\ref{tab1}. 
It is clear from the spectra shown in Fig.~\ref{fig9} and from 
Table~\ref{tab1} that at $T = 300$~K the lowest TO phonon is underdamped in the entire 
range of wave-vectors. Also the structure factors of the acoustic and optic branches 
do not change significantly with temperature as shown in Fig.~\ref{fig5}. 

The conclusion from this fitting is that although there is some temperature dependence 
of the phonon spectra at large wave-vectors q, at small wave vectors there is little 
temperature dependence and that we can essentially consider the low q high-energy phonons 
as temperature independent. 

\begin{figure}[h]
\centering
  \includegraphics[width=0.60\textwidth]{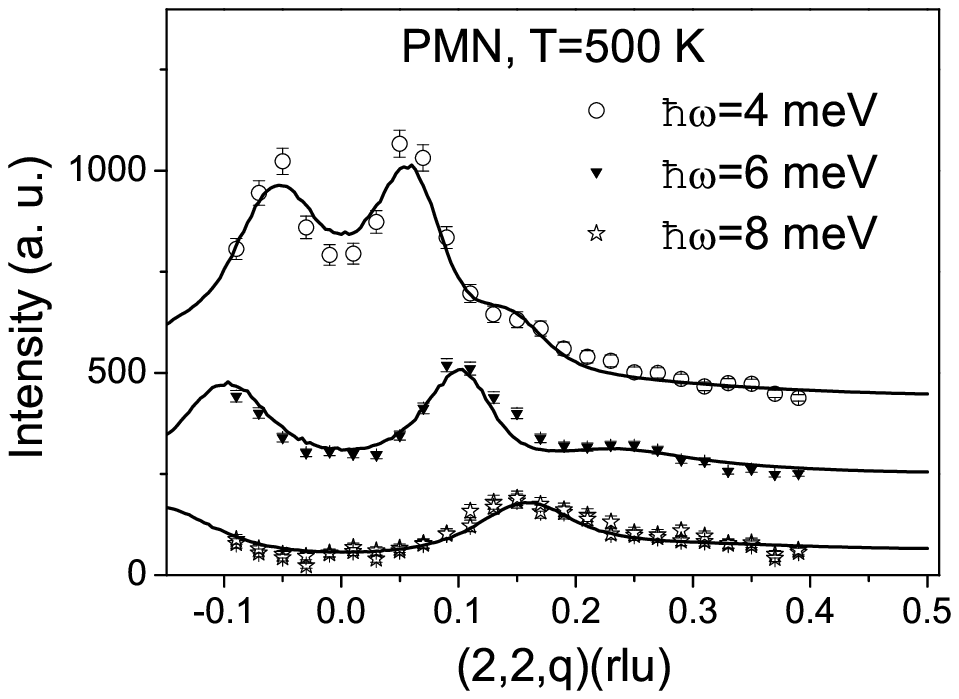}
  \caption{Constant-energy scans in PMN at $T=500$~K. For means of comparison
           the scans have been shifted by 200 counts.
           }
\label{fig10}
\end{figure}

The constant-energy scans in PMN along [2, 2, q] were measured to investigate the 
"waterfall effect" for $\hbar\omega$ = 4, 6 and 8~meV. The solid lines in Fig.~\ref{fig10} 
were obtained by varying only the intensities of the phonons and of the QE scattering 
while the other parameters of the phonons were fixed at the values obtained for $T = 300$~K. 
The model fits the experimental data very satisfactorily. We observe that the maxima in 
the spectra vary with the energy showing that the "waterfall" is not present at $T = 500$~K a 
least in (2,~2,~0) Brillouin zone.  
%
%
%
%
%
%
%
%
%
%
\subsection{Temperature dependence of the Elastic Scattering}
\label{tempelastic}
%
%
\begin{figure}[h]
\centering
  \includegraphics[width=0.60\textwidth]{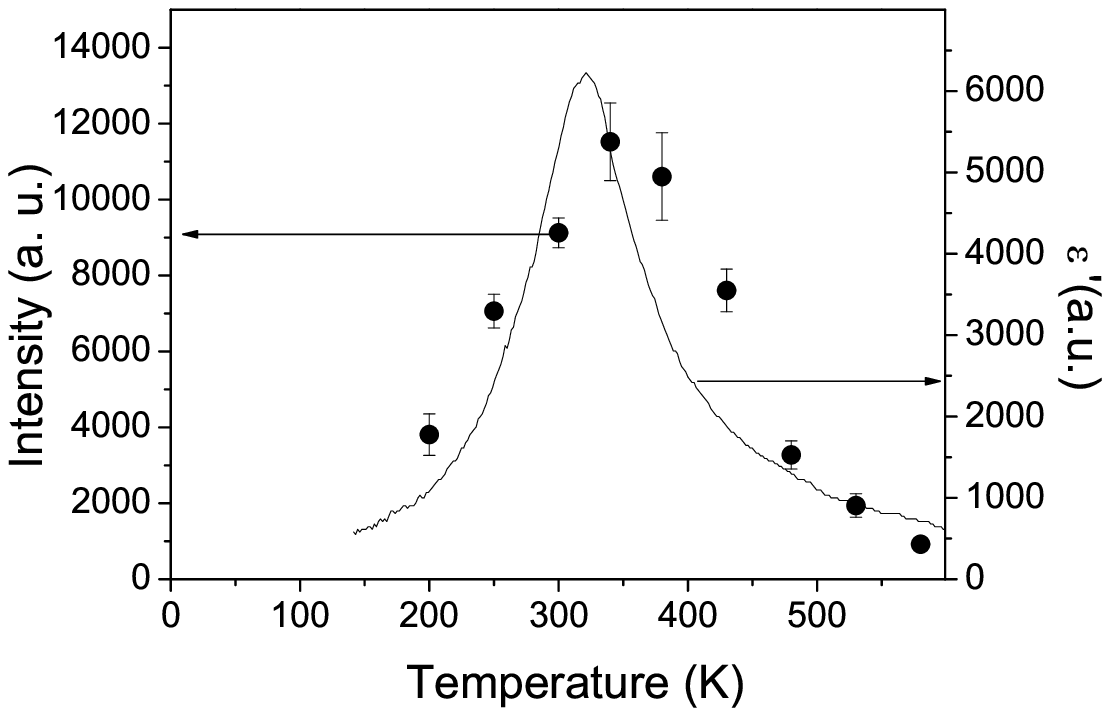}
  \caption{Temperature dependence of the susceptibility 
           of the QE scattering  in PMN measured 
           at $\mathbf{Q}=(1,~1,~0.05)$. For means of comparison the 
           real part of the dielectric permittivity of PMN at  
           frequency 1~GHz is also shown (Ref.~\cite{bovtun}).
           } 
\label{fig11}
\end{figure}
%
\begin{figure}[h]
\centering
  \includegraphics[width=0.60\textwidth]{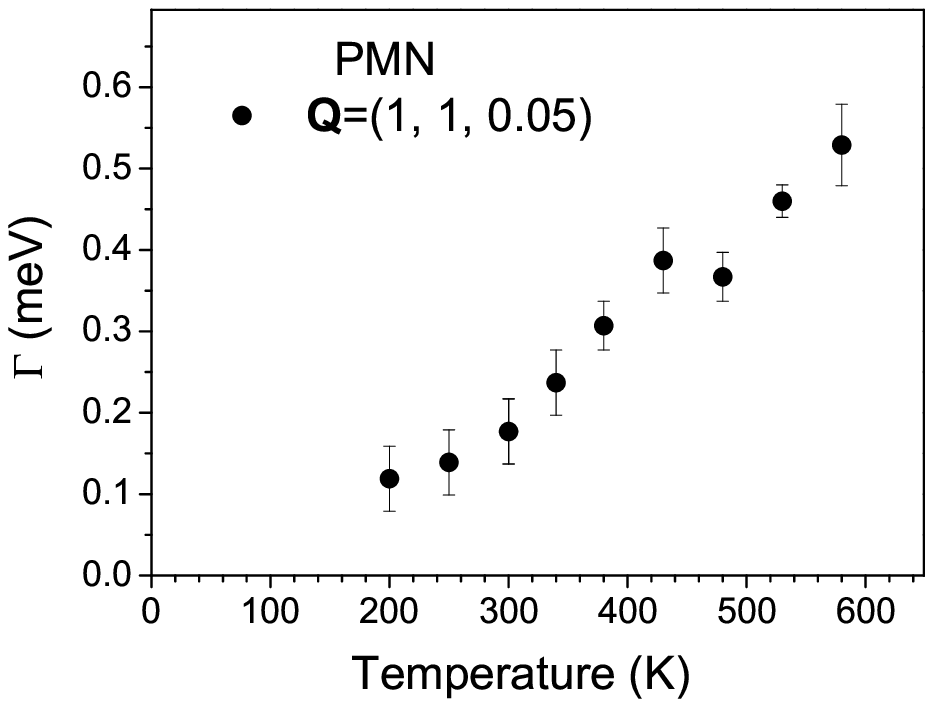}
  \caption{Temperature dependence of the damping of the 
           QE scattering measured at $\mathbf{Q}=(1,~1,~0.05)$. 
           }
\label{fig12}
\end{figure}
\begin{figure}[h]
\centering
  \includegraphics[width=0.60\textwidth]{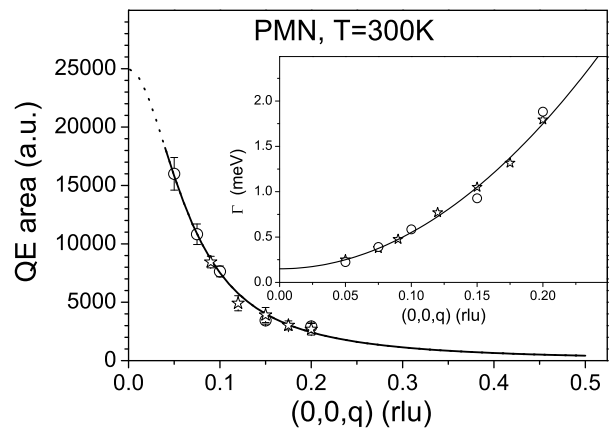}
  \caption{Wave-vector dependence of the energy-integrated intensity 
           of the QE scattering  in PMN measured in (1,~1,~0) (circles) 
           and (2,~2,~0) (stars) Brillouin zones. Intensities of the 
           QE component in two zones are scaled by a factor $\sim$1.3 
           at $q=0.2$~rlu. 
           The insert shows the 
           q-dependence of the damping of the QE component; 
           the error bars are within symbol size.
           }
\label{fig13}
\end{figure}
\begin{figure}[h]
\centering
  \includegraphics[width=0.60\textwidth]{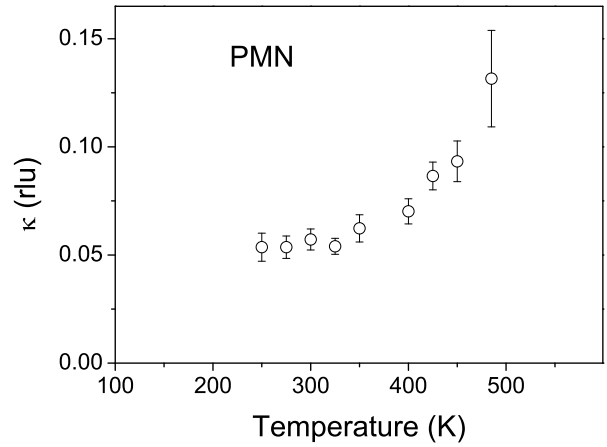}
  \caption{Temperature dependence of the inverse correlation length of 
           the quasi-elastic scattering.
           }
\label{fig14}
\end{figure}

%
The temperature dependence of the two components of the elastic scattering were also 
obtained from the fits of the model. Initially we shall consider the quasi-elastic 
scattering. The intensity of the quasi-elastic scattering is shown in Fig.~\ref{fig11}. 
It increases on cooling below 600 K reaching a maximum at about 370~K and then 
decreases on further cooling. The frequency width of the scattering is shown in 
Fig.~\ref{fig12} as deduced from measurements with $q = 0.05$~rlu. 
$\Gamma$ decreases with decreasing temperature from about 0.5~meV at 600~K to 
0.1~meV at 200~K. The wave-vector dependence of the width has been studied at 
temperatures of 450~K and 300~K and is shown in Fig.~\ref{fig13} for the 
latter temperature. It increases with wave-vector $q$ and is at least approximately 
proportional to a constant and a quadratic term, as suggested above by Eqn.~\ref{qe}. 
The wave-vector dependence of the energy integrated intensity of the quasi-elastic 
component is also shown in Fig.~\ref{fig13} and it is consistent with a Lorentzian 
line shape, as suggested in Eqn.~\ref{qe}. The values of the parameters characterizing 
the QE scattering at the temperatures of 450 K and 300 K are $\xi=7\pm2$~$\rm\AA$ 
and $11\pm2$~$\rm\AA$, $\Gamma_0=0.3\pm0.04$~meV and $0.15\pm0.03$~meV 
while $D_{QE}=19\pm2$~meV$^2\rm\AA^{2}$ and $17\pm2$~meV$^2\rm\AA^{2}$, respectively. 
The conclusion is that $D_{QE}$ does not change as a function of temperature, 
within the precision of the measurements, but that the inverse correlation length and 
the damping at $q=0$ both decrease as the temperature decreases and as illustrated 
for the inverse correlation length in Fig.~\ref{fig14}. Below 300 K, the correlation 
length remains approximately temperature independent but the damping still decreases.

\begin{figure}[h]
\centering
  \includegraphics[width=0.60\textwidth]{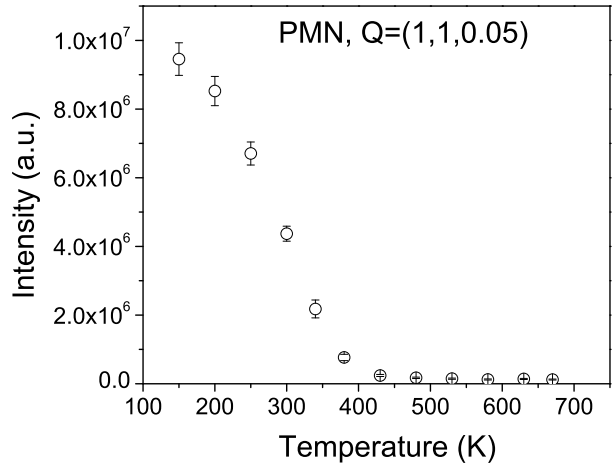}
  \caption{Temperature dependence of the intensity of the strictly 
           elastic central peak.
           }
\label{fig15}
\end{figure}

\begin{figure}[h]
\centering
  \includegraphics[width=0.60\textwidth]{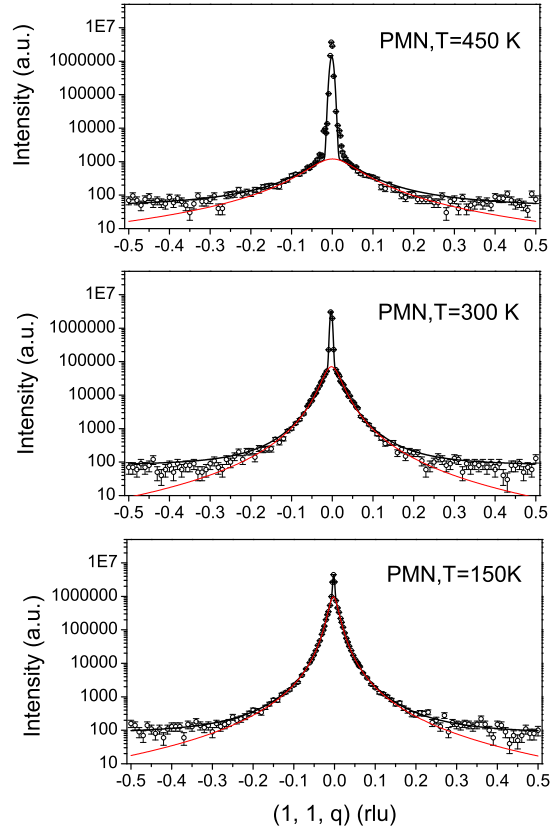}
  \caption{(Color online) Temperature dependence of the intensity of the 
           elastic scans.  These measurements were performed using $k_{f}=1.4$ 
           \AA$^{-1}$ and $10'/\rm \AA-20'-20'-80'$ collimation. 
           }
\label{fig16}
\end{figure}

The component of the central peak that is resolution limited is difficult to measure 
because it must be distinguished from the sharp and intense Bragg reflection on the 
one hand and from the incoherent scattering that is expected to vary only slowly with 
the wave-vector. Above about 480~K the intensity of the elastic scattering is weak 
and away from the Bragg reflection it probably arises from the incoherent scattering. 
Below a temperature of 370~K the scattering rapidly increases in intensity as the 
sample is cooled as shown in Fig.~\ref{fig15}. There appears to be some rounding 
possibly due to an inhomogeneous concentration distribution in the region 
of 370~K. The shape of the curve is nevertheless similar to that expected from 
an order parameter although in wave-vector it is wider than the Bragg reflection. 
The distribution of the intensity in wave-vector is determined by A($\mathbf Q$). 
It is known {\it e.g.} from reference~\cite{shirane20023} that the distribution of 
the diffuse scattering in PMN is largely perpendicular to the wave-vector transfer. 
In Fig.~\ref{fig16} we show the measured intensity perpendicular to the (1,~1,~0) 
Bragg reflection at 150~K, 300~K and 450~K. Clearly there is more diffuse scattering 
in the wings of the Bragg reflections at the two lower temperatures and in order to 
explain these results we have fitted the observations to the sum of a Bragg Gaussian, 
a flat background and a Lorentzian to an arbitrary power for the intermediate 
structure. As shown in Fig.~\ref{fig16} very reasonable descriptions were obtained 
when the intermediate peak was a Lorentzian to the power 1.5 while the inverse 
correlation function for this component slightly decreased from about 0.025(5) 
(rlu) at 300 K, to 0.015(5) (rlu) at 150 K. We note that at any temperature, 
the scattering by the central peak was found to be adequately described by a 
Lorenzian function to a power of about 1.5 and from our analysis of the data we conclude, 
that the correlation length decreases above room temperature to a value ~0.04 (rlu) 
at 400~K.  
%
%
%
%
%
%
%
%
\section{Discussion and Conclusion}
\label{concl}
\noindent 
\noindent 
We have measured the low energy phonons in the relaxor ferroelectric PMN 
using high resolution neutron scattering techniques. 
Previously~\cite{shirane20012,shirane20021,shirane20023,shirane20025} 
measurements were made of the neutron scattering in the (010) 
plane and using a more relaxed resolution function for the neutron scattering. 
Our results for the scattering in the (1$\bar{1}$0) plane lead to quite a 
different picture and so initially we will compare and contrast our 
experimental results. The important phonons in PMN are the TA and lowest 
TO branches and we agree that there is a strong coupling between these 
two modes of vibration and that they can strongly interact because they 
have the same symmetry. The earlier work found that above the Burns 
temperature 620~K the TO mode was underdamped and decreased in frequency 
as the temperature was decreased. Below the Burns temperature it became 
overdamped and the frequency could not be obtained until the sample was 
cooled below a temperature of 200~K when it increased in frequency and 
became underdamped again. Throughout this intermediate region between 200~K 
and 620~K the TO mode exhibited the "waterfall effect" at which the 
mode has an infinite slope and the wave-vector is almost constant. 
The critical wave-vector 0.2~$\rm \AA^{-1}$ suggested that this behaviour 
results from the interaction of the mode with the polar nano-regions (PNR) 
and with the strong coupling between the mode and the TA modes. Later 
neutron scattering experiments~\cite{kulda2} showed that the effect depends 
on the Brillouin zone and hence it is probably unlikely that it is due to 
the size of the PNR. 

Our results differ in that throughout the region immediately below the 
Burns temperature we find that the inelastic scattering can be described 
at least at small wave-vectors in terms of largely temperature independent 
underdamped modes particularly for the TO mode. We also did not observe a 
"waterfall effect" for the TO mode. We consider that the reason for the 
first difference is that we used substantially better resolution than 
the previous work while for the second difference our observations were 
made in a different Brillouin zone and so our results support the conclusion 
of Hlinka {\it et al.}~\cite{kulda2} based on their results for PZN. 

The high resolution of our results did enable us to distinguish two components 
at low energies; one was the quasi-elastic scattering and the other the central 
peak. We have measured the intensity scattered by the TO mode and by the 
QE scattering and found that the ratio differs from one Brillouin zone to another. 
For example, the TO mode is weak near (110) above the TA mode but the QE scattering 
is present. The QE scattering might arise from the TO mode through the mode 
coupling because at low frequencies, below the TA phonon the TO mode would have a 
different eigenvector from that above the TA mode. We would then expect a 
different structure factor for the TO mode and the QE scattering as discussed 
in reference ~\cite{shirane20023}. This effect does not however explain 
the difference in the intensities or the line shape of the scattering if the QE 
scattering is to be described by the interacting phonon model, outlined in the 
first part of section 3. We cannot however be certain that this is impossible if 
there is a strong frequency dependence to the self-energy of the TO mode, 
such as in the well known central peak problem in SrTiO$_3$~\cite{shirane1972}. 
Nevertheless we shall in this paper discuss the QE scattering as though it 
was an independent mode of the crystal.

We observe that the quasi-elastic scattering appears below the Burns temperature 
T$_d$. Upon cooling the sample the susceptibility of the QE scattering increases 
rapidly and peaks around $T_m=370$~K. Simultaneously the characteristic time of 
the fluctuations increases. The frequency and wave-vector dependence of the 
scattering are strongly suggestive of a phase transition at that temperature. 
Thus we can conclude that below T$_d$ we observe critical fluctuations that behave 
in a similar way to those of paraelectric or paramagnetic systems and one 
should expect a continuous phase transition when both the susceptibility and 
the correlation length diverge. The surprising feature 
is that the critical fluctuations in PMN have very low frequencies and 
are observed over an exceptionally wide temperature range between 200~K and 600~K. 
Most systems undergoing a phase transition have critical fluctuations that 
appear much closer to the phase transition and with much higher energies. This 
suggests that the fluctuations in PMN involve large collections of atoms, 
possibly PNR, rather than individual atoms. Below $T_m$ the susceptibility of the critical 
fluctuations in PMN decrease while the correlation length is approximately 
temperature independent equal to 11 $\rm\AA^{-1}$. This situation resembles the 
case of an antiferromagnet in a random-field.

The CP scattering is strictly elastic and so we shall assume that it is caused 
by short-range static displacements in which case it appears as diffuse scattering 
and not as Bragg scattering. The theory of the phase transition in an 
isotropic-three-dimensional ferroelectric predicts that the critical fluctuations 
will be perpendicular to the wave-vector transfer. This scattering is then similar 
to that expected if the critical fluctuations of an isotropic ferroelectric were 
frozen in at a temperature somewhat above the phase transition. We observe that the 
intensity of the elastic diffuse scattering (EDF) increases strongly below $T= 370$~K 
expected for the temperature dependence of an order parameter. The FWHM of the this 
scattering decreases below $T_m$ indicating that static displacements are correlated over 
larger distances. The diffuse scattering, however, never turns into a Bragg reflection. 
Above $T_m$ the diffuse scattering is very weak and disappears rapidly when approaching 
T$_d$  in agreement with the results of Hiraka {\it et al.}~\cite{shirane20042}. 
Above $T_m$, the scattering is much weaker and it was difficult to analyse the 
line shape precisely, due to the presence of the strong Bragg peak. Below $T_m$, the 
line-shape of the diffuse scattering is described by a Lorentzian function raised to 
the power of $z=1.5$. This is similar to the behaviour of an antiferromagnetic random 
field system in which the profile of the scattering is, after correction for 
the experimental resolution, a Lorentzian squared~\cite{shirane1982,shirane1989}. 
If the correction is not made the observed profile depends on the detail of the 
resolution function but in some cases is observed to be a Lorentzian to the power of 
about 1.5. This onset of the elastic EDF scattering in PMN is very suggestive of 
a random field transition at 370~K at which the ferroelectric critical scattering 
becomes pinned by the random field associated with the atomic disorder or PNR. 

It was suggested that the relaxors might be examples of random field transitions in 
reference~\cite{shirane20041}. However, a system with continuous symmetry is not 
expected to have a transition in the presence of a random field. Nevertheless 
PMN and other relaxors are cubic materials rather than being isotropic and the cubic 
symmetry breaks the continuous symmetry and allows for a random field transition. 
Most of the experimental work on random fields has been on random antiferromagnets in 
a uniform field which then generates a staggered antiferromagnetic 
field~\cite{shirane1982,shirane1989}. The experimental results then show that it is 
possible to reach a number of meta-stable states when the sample is cooled 
in a field. In the case of relaxors there has been little theoretical work 
developing a random field theory and in particular on the form of the transition 
in a cubic material. We suggest that as the sample is cooled the atomic disorder 
gives rise to dynamic clusters of PNR, as occurs in PMN below the Burn's temperature 
of 620~K. These do not order because the system is effectively isotropic at elevated 
temperature and fluctuations prevent the establishment of static order. At a temperature 
of about 370~K anisotropy starts to play a role and there is a cross-over to a 
cubic phase as opposed to an isotropic phase. The system then orders into a meta-stable 
random field phase that is stable down to the lowest temperature. The crystal reaches a 
uniform ferroelectric state if a large field is applied below a temperature of 210~K.

We are planning further experiments both to study the detailed shape of the ordering and 
to discover if we can observe any cross-over from isotopic to cubic symmetry as well as 
to study the QE scattering and the CP scattering in different Brillouin zones. 

\section{Acknowledgments}
This work was performed at the spallation neutron source SINQ, 
Paul Scherrer Institut, Villigen (Switzerland) and was partially 
supported by RFBR grant 05-02-17822.

\end{document}